\documentclass[a4paper,11pt]{article}

\usepackage[margin=1.14in]{geometry} 
\usepackage{tabularx}
\usepackage{multirow}
\usepackage{arydshln}
\usepackage{amsmath}
\usepackage{amsthm}
\usepackage{amssymb}
\usepackage{graphicx}
\usepackage{pdfpages}
\usepackage{enumitem}
\makeatletter
\makeatletter \renewcommand{\@dotsep}{10000} \makeatother
\usepackage{wrapfig}
\usepackage[utf8]{inputenc}
\usepackage{lmodern}
\usepackage{hyperref}
\hypersetup{
	unicode,
	colorlinks,
	breaklinks,
	urlcolor=cyan, 
    linkcolor=black, 
	pdfauthor={Author One, Author Two, Author Three},
	pdfproducer={LaTeX},
	pdfcreator={pdflatex}
}
\usepackage[T1]{fontenc} 
\usepackage{amsmath}

\newcommand{\be}{\begin{eqnarray}}
\newcommand{\ee}{\end{eqnarray}}
\def\be{\begin{equation}}
\def\ee{\end{equation}}
\def\bea{\begin{eqnarray}}
\def\eea{\end{eqnarray}}

\newcommand{\gsim}{\;\raisebox{-0.9ex}{$\textstyle\stackrel{\textstyle >}{\sim}$}\;}
\newcommand{\lsim}{\;\raisebox{-0.9ex}{$\textstyle\stackrel{\textstyle<}{\sim}$}\;}
\def\lsim{\raise0.3ex\hbox{$\;<$\kern-0.75em\raise-1.1ex\hbox{$\sim\;$}}}
\def\gsim{\raise0.3ex\hbox{$\;>$\kern-0.75em\raise-1.1ex\hbox{$\sim\;$}}}

\usepackage{graphics}

\usepackage{epsfig}
\usepackage{slashed}
\usepackage[utf8]{inputenc}
\usepackage{multirow}
\usepackage{pstricks}
\usepackage{dcolumn}

\usepackage[sort&compress,numbers,square]{natbib}

\usepackage{graphicx, color}
\usepackage[sort&compress,numbers,square]{natbib}
\def\ga{\mathrel{\raise.3ex\hbox{$>$\kern-.75em\lower1ex\hbox{$\sim$}}}}
\def\la{\mathrel{\raise.3ex\hbox{$<$\kern-.75em\lower1ex\hbox{$\sim$}}}}
\def\beqa{\begin{eqnarray}}
\def\eeqa{\end{eqnarray}}
\newcommand{\ov}{\overline}

\usepackage{graphicx, color}

\title{Exploring Exotic Decays of the Higgs Boson to Multi-Photons at the LHC via Multimodal Learning Approaches}
\vspace{8mm}
\author{\Large{A.\,Hammad$^{a}$, P. Ko$^{b}$,  Chih-Ting Lu$^{c}$ and Myeonghun Park$^{b,d}$}}

\date{
\small{
$^a$ Theory Center, IPNS, KEK, 1-1 Oho, Tsukuba, Ibaraki 305-0801, Japan.\\
$^b$ School of Physics, KIAS, Seoul 02455, Korea.\\
$^c$ Department of Physics and Institute of Theoretical Physics, Nanjing Normal University,
Nanjing, 210023, China.\\
$^d$ School of Natural Sciences, Seoultech, Seoul 01811, Korea.
 }
}
\begin{document}
	\maketitle
	\vspace{4mm}
	\begin{abstract}
The Standard Model (SM) Higgs boson, the most recently discovered elementary particle, may still serve as a mediator between the SM sector and a new physics sector related to dark matter (DM). The Large Hadron Collider (LHC) has not yet fully constrained the physics associated with the Higgs boson, leaving room for such possibilities. Among the various potential mass scales of the dark sector, the sub-GeV mass range is particularly intriguing. This parameter space presents significant challenges for DM direct detection experiments that rely on nuclear recoils. Various innovative experimental methods are currently under investigation to explore this sub-GeV dark sector. The LHC, functioning as a Higgs factory, could explore this sector once the challenge of identifying DM signals is resolved. Due to the significantly lower mass of particles in the dark sector compared to the Higgs boson, these particles are expected to be highly boosted following the Higgs boson's decay. However, detecting and identifying these highly boosted particles remains a considerable challenge at the LHC, despite their eventual decay into SM particles.  We employ a well-motivated leptophobic $Z^{\prime}_B$ model as a prototype to analyze the distinctive signatures from Higgs boson exotic decays into multi-photons. These signatures consist of collimated photons that fail to meet the photon isolation criteria, forming jet-like objects. Conventional analyses relying solely on the purity of energy deposits in the electromagnetic calorimeter would fail to detect these signatures, as they would be overwhelmed by background events from Quantum Chromodynamics. To effectively distinguish between such novel signal signatures and SM background events, we leverage advanced machine learning techniques, specifically the transformer encoder in a multimodal network structure. This neural network successfully separates dark sector signals from SM backgrounds, outperforming traditional event selection analyses.

\end{abstract}
\newpage
\noindent\rule{\textwidth}{1pt}
\tableofcontents
\noindent\rule{\textwidth}{0.2pt}
\maketitle \flushbottom
\vspace{4mm}
\section{Introduction}

The discovery of the Higgs boson\,\cite{Aad:2012tfa, Chatrchyan:2012ufa} provides strong evidence supporting the theory of electroweak symmetry breaking in the Standard Model of Particle Physics~\cite{Higgs:1964pj, Englert:1964et, Guralnik:1964eu}. However, unresolved puzzles remain beyond the Standard Model, including the origin of tiny neutrino masses, dark matter (DM), and matter-antimatter asymmetry in the universe. Dark sector models are a popular choice to address these issues, and the Higgs boson can act as a portal connecting the SM sector and the dark sector ~\cite{Patt:2006fw}\footnote{With the exception of the Higgs portal, the dark photon/dark Z portal presents an attractive approach to connect the visible sector with the dark sector~\cite{Fabbrichesi:2020wbt,Graham:2021ggy,Baek:2022ozm,Datta:2022zng,Ghosh:2023dgk}.}. In particular, if new particles are lighter than the Higgs boson, the exotic decays of the Higgs boson offer powerful ways to detect new physics beyond the SM ~\cite{Curtin:2013fra}.

Another mystery in the SM is the origin of accidental global $ U(1) $ symmetries of baryon and lepton numbers which may suggest the presence of new physics beyond the SM. A widely discussed proposal involves elevating these accidental global $ U(1) $ symmetries to gauged ones ~\cite{Foot:1989ts,FileviezPerez:2010gw,FileviezPerez:2011pt,Duerr:2013dza}. Furthermore, the $ U(1)_B $ gauge symmetry finds application in the $ E_6 $ grand unified theory~\cite{Babu:1996vt,Babu:1997st,Rizzo:1998ut}. Consequently, the leptophobic $ Z^{\prime}_B $ model emerges as one of the well-motivated $ U(1)^{\prime} $ models beyond the SM, particularly those incorporating DM candidates. Examples of UV completions for such models, with or without DM candidates, are detailed in Refs.~\cite{Ko:2010at,FileviezPerez:2010gw,Ko:2011di,FileviezPerez:2011pt,Duerr:2013dza,Ko:2016lai}. The scenario featuring light (baryonic) DM, in particular, stands out as one of the promising facets for DM models involving the leptophobic $ Z^{\prime}_B $~\cite{Ko:2011ns, Gondolo:2011eq}.
In this study, our focus is on a simplified model featuring $ U(1)_B $ gauge symmetry, assuming that both the leptophobic $ Z^{\prime}_B $ and the new scalar boson are much lighter than the Higgs boson. Specifically, the new scalar boson primarily decays into a pair of sub-GeV $ Z^{\prime}_B $ particles, with each $ Z^{\prime}_B $ predominantly decaying into a neutral pion and a photon, if kinematically allowed. 

The decay of the Higgs boson into a pair of $ Z^{\prime}_B $ or new scalar boson results in both of these light particles being highly boosted. Furthermore, the decay products from the highly boosted $ Z^{\prime}_B $ or new scalar boson tend to form a group of highly collimated photons. Notably, depending on the detector resolution and photon isolation criteria, this group of highly collimated photons can be identified as either a reconstructed photon ($\gamma_{\text{rec}}$) or a photon-jet ($J_{\gamma}$) \cite{Sheff:2020jyw, Lane:2023eno}.
If the opening angle distribution ($\Delta R_{\gamma\gamma}$) between a pair of photons within this group of collimated photons is smaller than the size of a single cell (i.e., $\Delta R_{\gamma\gamma} < 0.04$), the ATLAS detector cannot distinguish whether the signal originates from a single photon or a group of extremely collimated photons~\cite{ATLAS:2018dfo}. We refer to this signature as $\gamma_{\text{rec}}$. On the other hand, if some $\Delta R_{\gamma\gamma}$ distributions within the group of collimated photons fall within the range of $0.04 < \Delta R_{\gamma\gamma} < 0.4$, it is large enough to be distinguished from a single photon but too collimated to satisfy the photon isolation criteria. In such cases, we cannot identify them as multiple photons but rather as a jet-like structure of photons, denoted as $J_{\gamma}$, in the final state. 
The proposal for this kind of novel photon-jet signature was presented in Refs.~\cite{Dobrescu:2000jt, Toro:2012sv}. Investigating the structure of photon-jets involved the application of jet substructure techniques in Refs.~\cite{Ellis:2012sd, Ellis:2012zp, Sheff:2020jyw, Wang:2021uyb}, and machine learning techniques were employed in Refs.~\cite{Ren:2021prq, Wang:2023pqx, Ai:2024mkl}. Further studies on photon-jet signatures are available in Refs.~\cite{Bauer:2017nlg, Chakraborty:2017mbz, Sheff:2020jyw, Lu:2022zbe, Lane:2023eno, Alonso-Alvarez:2023wni}. Signal events of the Higgs boson exotic decays to multi-photons can classified into three categories, (1) two reconstructed photons~\cite{Draper:2012xt, ATLAS:2012soa, Bauer:2017nlg, Lane:2023eno}, (2) a reconstructed photon plus a photon-jet as the newly introduced signature here, and (3) two photon-jets which have been studied with jet substructures as above. 

As the signal with photon-jet faces huge QCD-related backgrounds, one needs to have additional handle more than an object tagging as previously studied. The phase space of the  signal events can be different from the one of background events, either due to the different structure of event-topology or a different weight from the amplitude. Thus, to maximize the performance in suppressing backgrounds, we need to optimize the process of  combining information from kinematics and QCD\,\cite{Ban:2023jfo, Hammad:2023sbd}.
Accordingly, for object tagging, we utilize an attention-based transformer encoder~\cite{vaswani2017attention} to distinguish the signal from the SM backgrounds. The proposed transformer model analyzes the events with the structure of particle clouds, which is inherited by the points cloud representation. Such a representation has the ability to share all the advantages of particle based representations, especially the flexibility to include arbitrary features for each particle. We endow this transformer model with extra tokens in a multimodal structure to facilitate the model ability to learn the full phase space of an event. Thus this network exploits the different kinematics of the signal from backgrounds. 

The paper is organized as follows. In Sec.~\ref{Sec:Model}, we introduce the simplified leptophobic $ Z^{\prime}_B $ model, while Sec.~\ref{Sec:Constraint} summarizes the relevant constraints for this model. Sec.~\ref{Sec:BPs} discusses the classification of signal signatures and presents benchmark points. Sec.~\ref{Sec:Simulation} showcases the signal-to-background analysis for Higgs boson exotic decays into a photon-jet and a reconstructed photon as well as two photon-jets at the LHC. All relevant  information for the transformer network used in this work is collected in Sec.~\ref{Sec:network}.  We present our results in Sec.~\ref{Sec:results}. Finally, our conclusions are presented in Sec.~\ref{Sec:Conclusion}.

\section{The simplified leptophobic $ Z^{\prime}_B $ model}\label{Sec:Model}

In this section, we initially introduce the simplified leptophobic $ Z^{\prime}_B $ model in Sec.~\ref{Sec:Lagrangian}. Subsequently, we delineate the exotic partial decay widths of the SM-like Higgs boson $ h_2 $ in Sec.~\ref{Sec:h2decay}. The partial decay widths of the new light scalar boson $ h_1 $ and $ Z^{\prime}_B $ can be found in the Appendix~\ref{Sec:h1_decay} and Appendix~\ref{Sec:Zp_decay}, respectively.
\subsection{The model}\label{Sec:Lagrangian}

The model with $U(1)_B$ gauge symmetry is one of the well-motivated $U(1)^{\prime}$ models beyond the SM~\cite{Ko:2010at,FileviezPerez:2010gw,Ko:2011di,FileviezPerez:2011pt,Duerr:2013dza,Ko:2016lai}. In this simplified model, we assume that all exotic new particles (exotic fermions for anomaly cancellations and
extra scalar bosons that provide masses to the exotic fermions and/or make exotic charged/colored particles decay fast enough) are at a relatively high energy scale, except for the light leptophobic $Z^{\prime}_B$ and new scalar boson $h_1$. The associated $U(1)_B$ charges $Q^i_B$ for fields in our simplified model are listed in Table~\ref{U1coup}\footnote{Note that the Dirac (Majorana) fermion DM candidate $\chi_D$ ($\chi_M$) with $Q^i_B(\chi_{D,M})=1$ can be involved in this simplified leptophobic $Z^{\prime}_B$ model. Since the DM part is not the focus of this study, we will address this issue in the future.}. 

\begin{table}[th!]
\centering
  \begin{tabular}{l @{\extracolsep{0.2in}} r r r r}
   \hline
Field & quarks & leptons & $ H $ & $ \Phi $ \\ \hline
$ Q^i_B $ & $1/3$ & $ 0 $ & $ 0 $ & $ -2 $ \\ \hline
    \end{tabular}
    \caption{\small
    The associated $ U(1)_B $ charges $ Q^i_B $ for fields in our simplified leptophobic $ Z^{\prime}_B $ model.}
\label{U1coup}
\end{table}

Here, $H$ represents the SM Higgs doublet, and $\Phi$ is a new complex SM singlet scalar responsible for generating the mass term for $Z^{\prime}_B$ via the Higgs mechanism.

The Lagrangian in this simplified model can be represented as
\begin{equation}
{\cal L}={\cal L}_{\text{SM}}+{\cal L}^{\prime}_{\text{scalar}}+{\cal L}^{\prime}_{\text{gauge}}, 
\end{equation}
where
\begin{align*}
& {\cal L}'_{\text{scalar}} =
D_\mu \Phi^\dagger D^\mu \Phi 
- m_\Phi^2 \Phi^\dagger \Phi 
- \lambda_\Phi (\Phi^\dagger \Phi)^2
- \lambda_{H\Phi} H^\dagger H\Phi^\dagger \Phi,
\\ &
{\cal L}'_{\text{gauge}} =
- \frac{1}{4}  X_{\mu\nu}X^{\mu\nu}
- \frac{\epsilon}{2}  B_{\mu\nu}X^{\mu\nu}
+ g_{B}\sum_{f}Q^f_B \ov{f}\rlap{\,/}X f.
\end{align*}
The ${\cal L}_{\text{SM}}$ is the full SM Lagrangian and $f$ stands for SM fermions. Covariant derivative can be represented as
\begin{equation}
D_\mu = \partial_\mu - i g_B Q^i_B X_\mu.
\end{equation}
 $ \epsilon $ is the kinetic mixing parameter betweenthe two $U(1)$s, $g_B$ is the $U(1)_B$ gauge coupling constant and $Q^i_B$ is shown in Table~\ref{U1coup}.

Expanding the scalar fields around the vacuum expectation values (VEVs) with the unitary gauge as 
\begin{equation}
H(x) = \frac{1}{\sqrt 2} 
\left(
\begin{tabular}{c}
0
\\
$v + h(x)$
\end{tabular}
\right)
\;\;\; , \;\;\; 
\Phi (x) = \frac{1}{\sqrt 2} \left( v_B + h_B (x) \right). 
\end{equation}
After applying the condition for potential minimization, the VEVs $ v $ and $ v_B $ can be written as
\begin{equation}
v^2 = \frac{4\lambda_{\Phi}\mu^2 -2\lambda_{H\Phi}\mu^2_{\Phi}}{4\lambda_{\Phi}\lambda -\lambda^2_{H\Phi}},\quad v^2_B = \frac{4\lambda\mu^2_{\Phi} -2\lambda_{H\Phi}\mu^2}{4\lambda_{\Phi}\lambda -\lambda^2_{H\Phi}}, 
\end{equation} 
where $\lambda$ is the quartic self-coupling. Finally, these two neutral scalar bosons $ h(x) $ and $ h_B(x) $ will mix to each other after the symmetry breaking, and they can be diagonalized to the mass eigenstates $ h_1, h_2 $ via
\begin{equation}
\tan 2\theta_h = \frac{\lambda_{H\Phi}vv_B}{\lambda_{H\Phi}v^2_B -\lambda v^2}, 
\end{equation}
and 
\begin{equation}
M^2_{h_1,h_2}=(\lambda v^2 +\lambda_{H\Phi}v^2_B)\mp\sqrt{(\lambda v^2 -\lambda_{H\Phi}v^2_B)^2 +\lambda_{H\Phi}v^2 v^2_B}.
\end{equation}
In our convention,  we assign $h_2$ as the SM-like Higgs boson with $M_{h_2} = 125$ GeV, and $h_1$ as the light scalar boson with $M_{h_1} \ll M_{h_2}$.

On the other hand, the three neutral gauge bosons $ B_{\mu}, W^3_{\mu}, X_{\mu} $ are diagonalized to the mass eigenstates $ A_{\mu}, Z_{\mu}, Z^{\prime}_{B\mu} $\footnote{
The details of this digonalization can be found in Ref.~\cite{Feldman:2007wj,Wells:2008xg,Chang:2013lfa} and we will not repeat them again.
},
\begin{equation}
M^2_{\gamma}=0,\quad M^2_{Z,Z^{\prime}_B}=(\Delta_1 +\Delta_2)/2,  
\end{equation}
where 
\begin{align*}
& \Delta_1 =
g^2_B (Q^{\Phi}_B)^2 v^2_B \eta^2 + \frac{1}{4}(g^2 +g'^2 \eta^2)v^2, \hspace{4mm}
\Delta_2 = \sqrt{\Delta_1^2 -g^2_B (Q^{\Phi}_B)^2 v^2 v^2_B (g^2 +g'^2)\eta^2}.
\end{align*}
We assume $ M_{Z^{\prime}_B}\ll M_Z $ and $ \eta\equiv\epsilon /\sqrt{1-\epsilon^2} $. Considering  small kinematic mixing $ \epsilon $, as in this study, $ M_Z $ and $ M_{Z^{\prime}_B} $ can be approximated as
\begin{equation}
M_Z\simeq \sqrt{g^2 +g'^2}v/2,\quad M_{Z^{\prime}_B}\simeq g_B |Q^{\Phi}_B| v_B. 
\end{equation}

Although SM charged leptons do not carry the $U(1)_B$ charge, the $Z^{\prime}_B$ cannot be completely decoupled from them. The kinematic mixing between $Z^{\prime}_B$ and the photon, resulting from one-loop radiative corrections involving SM quarks and heavy charged fermions, can generate a parameter $\epsilon \sim \frac{eg_B}{(4\pi)}$~\cite{Carone:1995pu,Aranda:1998fr}. We will consider this effect in our calculations below.

\subsection{Exotic Higgs boson decays}\label{Sec:h2decay}

The light scalar boson $h_1$ and the leptophobic $Z^{\prime}_B$ can lead to interesting exotic decays of the SM-like Higgs boson, $h_2$, via the mixing between $H$ and $\Phi$. Depending on the mass spectrum of $h_1$ and $Z^{\prime}_B$, there are at least four possibilities for the exotic decays of $h_2$:
\begin{equation}
h_2\rightarrow Z^{\prime}_B Z^{\prime}_B,\quad h_2\rightarrow h_1 h_1,\quad h_2\rightarrow h_1 h^{\ast}_1\rightarrow h_1 Z^{\prime}_B Z^{\prime}_B,\quad h_2\rightarrow h_1 h_1 h_1.
\end{equation} 
We are aware that the three-body decays in the third and fourth cases suffer from phase space suppression. Although these two decay channels can be numerically calculated, as shown in Ref.~\cite{Chang:2013lfa}. As their contributions are much smaller compared to the first two decay channels, we will only consider the partial decay widths for the first two cases below.

The decay width of $ h_2 $ can be represented as
\begin{equation}
\Gamma_{h_2} = \cos^2\theta_h\widehat{\Gamma}_{h_2}+\Gamma^{\prime}_{h_2}, 
\label{eq:h2width-1}
\end{equation}
where $ \widehat{\Gamma}_{h_2} $ is the  SM Higgs boson width, $ 4.03 $ MeV~\cite{LHCHiggsCrossSectionWorkingGroup:2011wcg}, and
\begin{equation}
\Gamma^{\prime}_{h_2}\approx \sin^2\theta_h\widehat{\Gamma}(h_2\rightarrow Z^{\prime}_B Z^{\prime}_B)+\Gamma (h_2\rightarrow h_1 h_1), 
\label{eq:h2width-2}
\end{equation}
where
\begin{align}
& \widehat{\Gamma}(h_2\rightarrow Z^{\prime}_B Z^{\prime}_B) = \frac{g^2_B M^2_{Z^{\prime}_B}}{8\pi M_{h_2}}\sqrt{1-\frac{4M^2_{Z^{\prime}_B}}{M^2_{h_2}}}(3-\frac{M^2_{h_2}}{M^2_{Z^{\prime}_B}}+\frac{M^4_{h_2}}{4M^4_{Z^{\prime}_B}}), \label{eq:h2ZpBZpB_width} \\ &
\Gamma (h_2\rightarrow h_1 h_1) = \frac{\mu^2_{h_2 h_1 h_1}}{32\pi M_{h_2}}\sqrt{1-\frac{4M^2_{h_1}}{M^2_{h_2}}}, 
\end{align}
with 
\begin{align*}
\mu_{h_2 h_1 h_1} = & 2\cos^3 \theta_h \lambda_{H\Phi}v +2\sqrt{2}\cos^2 \theta_h \sin\theta_h (\lambda_{H\Phi}-6\lambda_{\Phi})v_B
\nonumber \\ &
+ \cos\theta_h \sin^2\theta_h (6\lambda -4\lambda_{H\Phi})v -\sqrt{2}\sin^3 \theta_h \lambda_{H\Phi} v_B.
\end{align*}
Assuming that $ 2 M_{Z^{\prime}_B} < M_{h_1} \ll M_{h_2} $ and fix $ \alpha_B = 10^{-5} $, $ M_{Z^{\prime}_B} = 0.4 $ GeV and $ \sin\theta_h = 1.6\times 10^{-4} $, the branching ratios of $ h_2\rightarrow Z^{\prime}_B Z^{\prime}_B $ and $ h_2\rightarrow h_1 h_1 $ can be as large as $ 4.35\times 10^{-4} $ and $ 8.70\times 10^{-4} $, respectively.
\begin{figure}[t!]\centering
\includegraphics[width=0.48\textwidth]{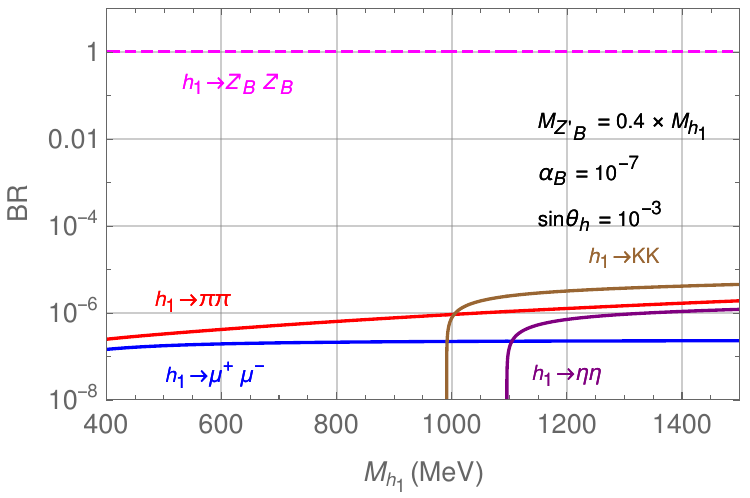}
\includegraphics[width=0.48\textwidth]{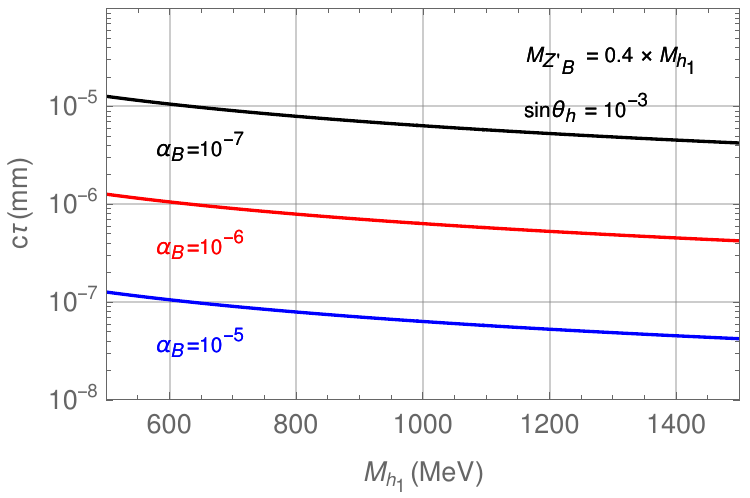}
\caption{
The branching ratios of $ h_1 $ when $ h_1\rightarrow Z^{\prime}_B Z^{\prime}_B $ is kinematically allowed with fixed $ \alpha_B = 10^{-7} $, $ \sin\theta_h = 10^{-3} $, and $ M_{Z^{\prime}_B}=0.4 M_{h_1} $ (left panel) and the proper decay length ($ c\tau $) [mm] of $ h_1 $ with fixed $ \sin\theta_h = 10^{-3} $, $ M_{Z^{\prime}_B}=0.4 M_{h_1} $ and varying $ \alpha_B = 10^{-5} $, $ 10^{-6} $ and $ 10^{-7} $ (right panel).
}
\label{fig:MhDBR2}
\end{figure}
The lighter scalar, $ h_1 $, can couple either directly to the $ Z^{\prime}_B $ pair or to SM particles via the mixing between $ H $ and $ \Phi $. Analytical formulas for the  partial decay widths of $ h_1 $ can be found in Appendix~\ref{Sec:h1_decay}. 

In this study, we focus on the case where$ h_1 \rightarrow Z^{\prime}_B Z^{\prime}_B $ is kinematically allowed. We fix $ \alpha_B = 10^{-7} $, $ \sin\theta_h = 10^{-3} $, and $ M_{Z^{\prime}_B} = 0.4 M_{h_1} $ to illustrate the branching ratios of $ h_1 $ in the left panel of Fig.~\ref{fig:MhDBR2}. The right panel of Fig.~\ref{fig:MhDBR2} displays the proper decay lengths ($ c\tau $) [mm] of $ h_1 $ with varying $ \alpha_B = 10^{-5} $, $ 10^{-6} $, and $ 10^{-7} $ while keeping $ \sin\theta_h = 10^{-3} $ and $ M_{Z^{\prime}_B} = 0.4 M_{h_1} $ fixed. We observe that once $ h_1 \rightarrow Z^{\prime}_B Z^{\prime}_B $ is kinematically allowed, it becomes the dominant decay mode, and $ h_1 $ promptly decays to the final state for the entire parameter space.

\begin{figure}[t!]\centering
\includegraphics[width=0.48\textwidth]{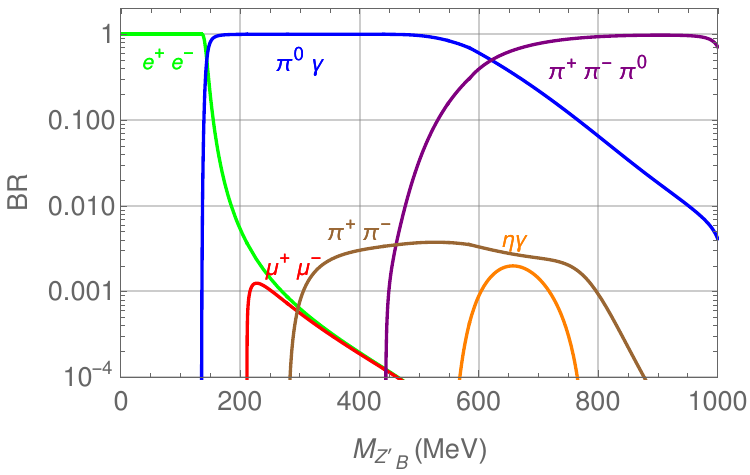}
\includegraphics[width=0.48\textwidth]{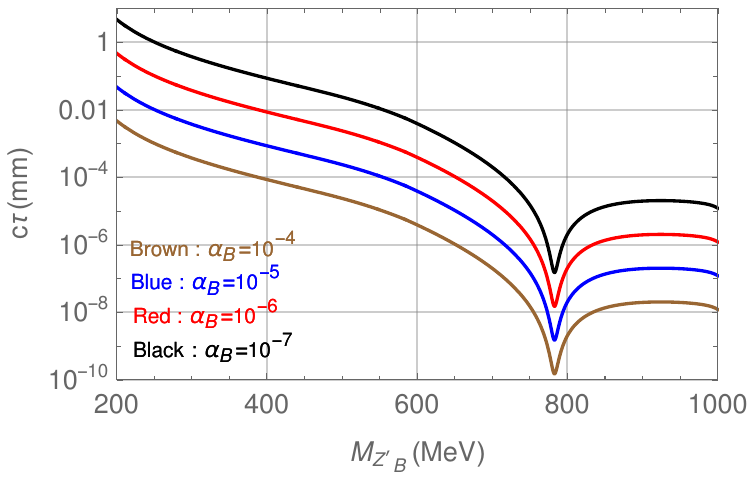}
\caption{
The decay branching ratios of sub-GeV $ Z^{\prime}_B $ (left panel) and its proper decay length ($ c\tau $) [mm] with $ \alpha_B = 10^{-4} $, $ 10^{-5} $, $ 10^{-6} $ and $ 10^{-7} $ (right panel). Here we assume $ \epsilon = 0.1\times\frac{eg_B}{(4\pi)} $.
}
\label{fig:ZpBBR}
\end{figure}

The important feature of the light $ Z^{\prime}_B $ is its decay branching ratios, which differ significantly from the usual kinematic mixing dark photon scenario. For $ M_{Z^{\prime}_B} \lesssim 1 $ GeV, the dominant decay modes of $ Z^{\prime}_B $ are $ Z^{\prime}_B \rightarrow \pi^0\gamma $ and $ Z^{\prime}_B \rightarrow \pi^+\pi^-\pi^0 $. However, $ Z^{\prime}_B \rightarrow e^+ e^- $ becomes the dominant decay mode when pion decays are kinematically forbidden. The partial decay widths for $ Z^{\prime}_B $ to light mesons are calculated using the vector meson dominance (VMD) model~\cite{Sakurai:1960ju}, and the detailed partial decay widths of $ Z^{\prime}_B $ are shown in Appendix~\ref{Sec:Zp_decay}, following closely Ref.~\cite{Tulin:2014tya}. In the left panel of Fig.~\ref{fig:ZpBBR}, we present the decay branching ratios for sub-GeV $ Z^{\prime}_B $. The right panel of Fig.~\ref{fig:ZpBBR} displays the proper decay length ($ c\tau $) [mm] of $ Z^{\prime}_B $ with different values of $ \alpha_B = 10^{-4} $, $ 10^{-5} $, $ 10^{-6} $, and $ 10^{-7} $. In order to avoid constraints from KLOE and LHCb, which depend on the couplings of $ Z^{\prime}_B $ to lepton pairs, we assume $ \epsilon = 0.1\times\frac{eg_B}{(4\pi)} $ in Fig.~\ref{fig:ZpBBR}. Importantly,  the light $ Z^{\prime}_B $ can decay promptly when $ \pi^0\gamma $ is kinematically allowed\footnote{For the case of a light $ Z^{\prime}_B $ as a long-lived particle (LLP) with a small enough $ \alpha_B $ at the LHC, the charged track information from the displaced vertex in the tracker system can record the decay mode of $Z^{\prime}_B\rightarrow\pi^+\pi^-\pi^0$. However, effectively finding a trigger for the decay mode of $Z^{\prime}_B\rightarrow\pi^0\gamma$ when the $Z^{\prime}_B$ becomes the LLP poses a challenge.}.

\section{Constraining the model free parameters }\label{Sec:Constraint}
In this section, we consider various constraints in the simplified leptophobic $ Z^{\prime}_B $ model. First, we address the constraints on $h_2$ and $h_1$. The global fits of Higgs boson data from ATLAS in Ref.~\cite{ATLAS:2020qdt} suggest that the allowed branching ratio of $h_2$ exotic decays is below $19\%$ at the $95\%$ confidence level. We apply this constraint to our scenario as follows:
\begin{equation}
\frac{\Gamma^{\prime}_{h_2}}{\Gamma_{h_2}} \lesssim 0.19,
\label{eq:Higgs_exotic}
\end{equation}
where $ \Gamma_{h_2} $ and $ \Gamma^{\prime}_{h_2} $ are given in Eqs.~(\ref{eq:h2width-1}) and~(\ref{eq:h2width-2}), respectively. Additionally, $h_1$ with $ M_{h_1} \lesssim {\cal O}(1) $ GeV is subject to severe constraints from fixed-target experiments and $ B $ meson decay processes if $ h_1\rightarrow Z^{\prime}_B Z^{\prime}_B $ is kinematically forbidden. Regions of $ M_{h_1} < 280 $ MeV with $ \sin\theta_h \geq 10^{-5} $ have already been ruled out. However, there are still some allowed parameter space with sizable $ \sin\theta_h $ for $ M_{h_1} > 280 $ MeV. Since we focus on $ M_{h_1} \geqslant 1 $ GeV and $ h_1\rightarrow Z^{\prime}_B Z^{\prime}_B $ becomes the dominant decay channel in this work, the above constraints are much weaker in our scenario. Therefore, we conservatively require $ \sin\theta_h\lesssim 10^{-3}$ to the parameter space of interest. For more details on constraints from a light scalar mixed with the SM Higgs boson, refer to Refs.~\cite{Schmidt-Hoberg:2013hba,Clarke:2013aya,Chang:2016lfq}.

\begin{figure}[t!]\centering
\includegraphics[width=0.8\textwidth]{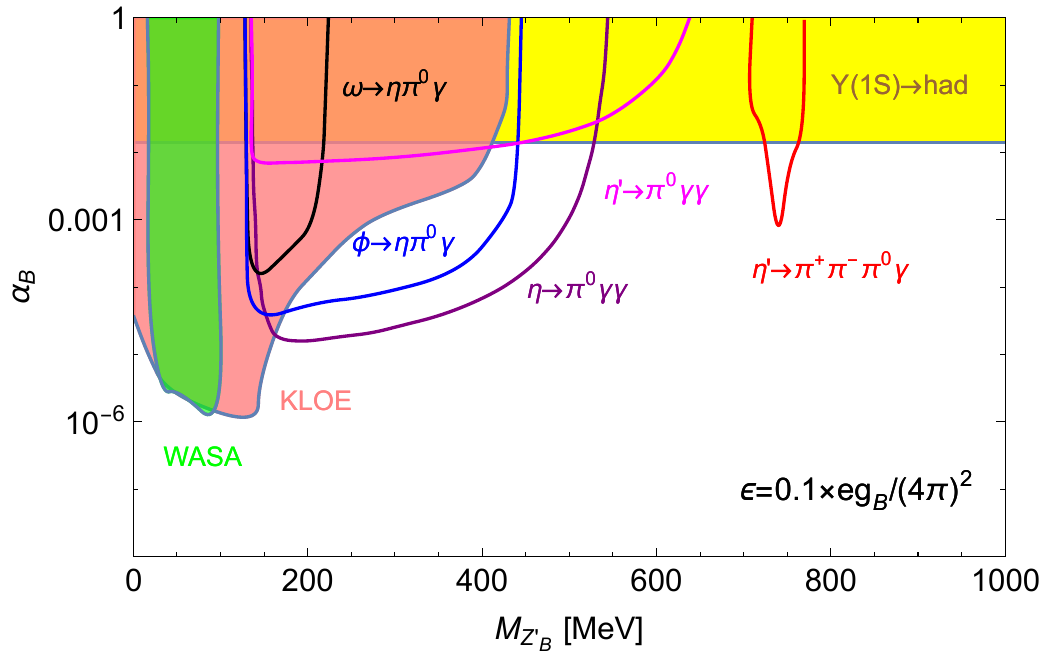}
\caption{
Various constraints for the light leptophobic $ Z^{\prime}_B $ on the $(M_{Z^{\prime}_B},\alpha_B)$ plane. Each constraint is explained in the main text.
}
\label{fig:ZpBcons}
\end{figure}

The sub-GeV leptophobic $ Z^{\prime}_B $ is subject to severe constraints from various meson decay processes. We will begin by summarizing potential constraints from the $ Z^{\prime}_B\rightarrow\pi^0 \gamma $ decay channel:
\begin{itemize}
\item $ \eta\rightarrow\pi^0 \gamma\gamma $: 
An updated measurement of $ \eta\rightarrow\pi^0 \gamma\gamma $ was published by Crystal-Ball@MAMI~\cite{Nefkens:2014zlt}, yielding a more precise result for the partial decay width: $ \Gamma (\eta\rightarrow\pi^0 \gamma\gamma) = (0.33\pm 0.03_{\text{tot}}) $ eV. A previous reanalysis of data from CrystalBall@AGS~\cite{Prakhov:2008zz} also provided similar results with $ \Gamma (\eta\rightarrow\pi^0 \gamma\gamma) = (0.285\pm 0.068_{\text{tot}}) $ eV. Here we applied the value reported by the PDG, BR$ (\eta\rightarrow\pi^0 \gamma\gamma) = (2.56\pm 0.22)\times 10^{-4} $~\cite{ParticleDataGroup:2020ssz,Escribano:2022njt}. This measurement can be applied to constrain the branching ratio, BR$ (\eta\rightarrow Z^{\prime}_B\gamma\rightarrow\pi^0 \gamma\gamma) $, for the $ Z^{\prime}_B $, as indicated by the purple line in Fig.~\ref{fig:ZpBcons}.
\item $ \eta^{\prime}\rightarrow\pi^0 \gamma\gamma $: 
A new measurement of $ \eta^{\prime}\rightarrow\pi^0 \gamma\gamma $ was reported by BESIII~\cite{Ablikim:2016tuo}, yielding a more precise result for the nonresonant branching fraction: BR$ (\eta^{\prime}\rightarrow\pi^0 \gamma\gamma) = (6.16\pm 0.64(\text{stat})\pm 0.67(\text{sys}))\times 10^{-4} $. This measurement can be used to constrain the branching ratio BR$ (\eta^{\prime}\rightarrow Z^{\prime}_B\gamma\rightarrow\pi^0 \gamma\gamma) $ for the $ Z^{\prime}_B $, as illustrated by the magenta line in Fig.~\ref{fig:ZpBcons}.
\item $ \phi\rightarrow\eta\pi^0 \gamma $: 
The measurement of $ \phi\rightarrow\eta\pi^0 \gamma $ by KLOE~\cite{Ambrosino:2009py} provides a constraint on the branching fraction, BR$ (\phi\rightarrow\eta\pi^0 \gamma) = (7.06\pm 0.22)\times 10^{-5} $. This constraint can be applied to the branching ratio BR$ (\phi\rightarrow\eta Z^{\prime}_B\rightarrow\eta\pi^0 \gamma) $ for the $ Z^{\prime}_B $, as illustrated by the blue line in Fig.~\ref{fig:ZpBcons}.
\item $ \omega\rightarrow\eta\pi^0 \gamma $: 
The upper limit on the branching fraction, Br$ (\omega\rightarrow\eta\pi^0 \gamma) < 3.3\times 10^{-5} $ ($90\%$ C.L.), was obtained from CMD-2@VEPP-2M~\cite{Akhmetshin:2003rg}. This constraint can be applied to the branching ratio BR$ (\omega\rightarrow\eta Z^{\prime}_B\rightarrow\eta\pi^0 \gamma) $ for the $ Z^{\prime}_B $, as shown by the black line in Fig.~\ref{fig:ZpBcons}. 
\end{itemize} 
Next, we consider the constraint from the $ Z^{\prime}_B\rightarrow\pi^+ \pi^- \pi^0 $ decay mode with $ M_{Z^{\prime}_B} > 620 $ MeV within the $ \eta^{\prime}\rightarrow\pi^+ \pi^- \pi^0 \gamma $ channel. CLEO conducted a search for $ \eta^{\prime}\rightarrow\omega\gamma\rightarrow\pi^+ \pi^- \pi^0 \gamma $ from charmonium decays with BR$ (\eta^{\prime}\rightarrow\omega\gamma) = (2.34\pm 0.30\pm 0.04)\% $~\cite{Pedlar:2009aa}. However, BESIII updated the measurement to BR$ (\eta^{\prime}\rightarrow\omega\gamma) = (2.489\pm 0.018\pm 0.074)\% $~\cite{BESIII:2019gef}. We will apply the BESIII measurement to constrain BR$ (\eta^{\prime}\rightarrow Z^{\prime}_B\gamma\rightarrow\pi^+ \pi^- \pi^0 \gamma) $ for the $ Z^{\prime}_B $, as indicated by the red line in Fig.~\ref{fig:ZpBcons}.

Notice some constraints from KLOE~\cite{Babusci:2012cr} and WASA~\cite{Adlarson:2013eza} are dependent on couplings of $ Z^{\prime}_B $ to leptons which are proportional to $ \epsilon $. Here we assume $ \epsilon $ is much smaller ($ \epsilon < 0.1\times\frac{eg_B}{(4\pi)^2} $) in this study, so those constraints become much weaker~\cite{Tulin:2014tya,Ilten:2018crw} as shown for the pink and green bulks in Fig.~\ref{fig:ZpBcons}. 
Finally, the hadronic decay of $ \Upsilon (1S) $ state mediated by an s-channel $ Z^{\prime}_B $ provide $ \alpha_B < 0.014 $ which is independent of $ M_{Z^{\prime}_B} $ if $ \sqrt{s}\gg M_{Z^{\prime}_B} $~\cite{Aranda:1998fr} as shown for the yellow bulk in Fig.~\ref{fig:ZpBcons}. Therefore, $\alpha_B\lesssim 10^{-5}$ is still allowed for $M_{Z^{\prime}_B}\gtrsim 200$ MeV.

\section{Search strategy}\label{Sec:BPs}

Among the various processes offered by the rich parameter space of our benchmark leptophobic $ Z^{\prime}_B $ model, we explore two intriguing topologies arising from Higgs boson exotic decays into multiple photons in the Higgs production through gluon fusion:
\begin{align}
& \text{TP1}: g g\rightarrow h_2\rightarrow Z^{\prime}_B Z^{\prime}_B\rightarrow (\pi^0 \gamma)(\pi^0 \gamma)\rightarrow 6\gamma , 
\nonumber \\
& \text{TP2}: g g\rightarrow h_2\rightarrow h_1 h_1\rightarrow (Z^{\prime}_B Z^{\prime}_B) (Z^{\prime}_B Z^{\prime}_B)\rightarrow (\pi^0 \gamma)(\pi^0 \gamma)(\pi^0 \gamma)(\pi^0 \gamma)\rightarrow 12\gamma .\label{eq:process}
\end{align} 
Since we consider  $ M_{h_1}\sim {\cal O}(1) $ GeV and a sub-GeV $ Z^{\prime}_B $, both $ h_1 $ and $ Z^{\prime}_B $ are highly boosted. Photons emerging  from highly boosted $ h_1 $ and $ Z^{\prime}_B $ are collimated enough and can be identified as a single photon-like energy cluster, $\gamma_{\text{rec}}$, or a jet-like structure, $J_{\gamma}$, in the electromagnetic  calorimeter depending on the opening angles among these photons.

\begin{figure}[h!]\centering
\includegraphics[width=0.9\textwidth]{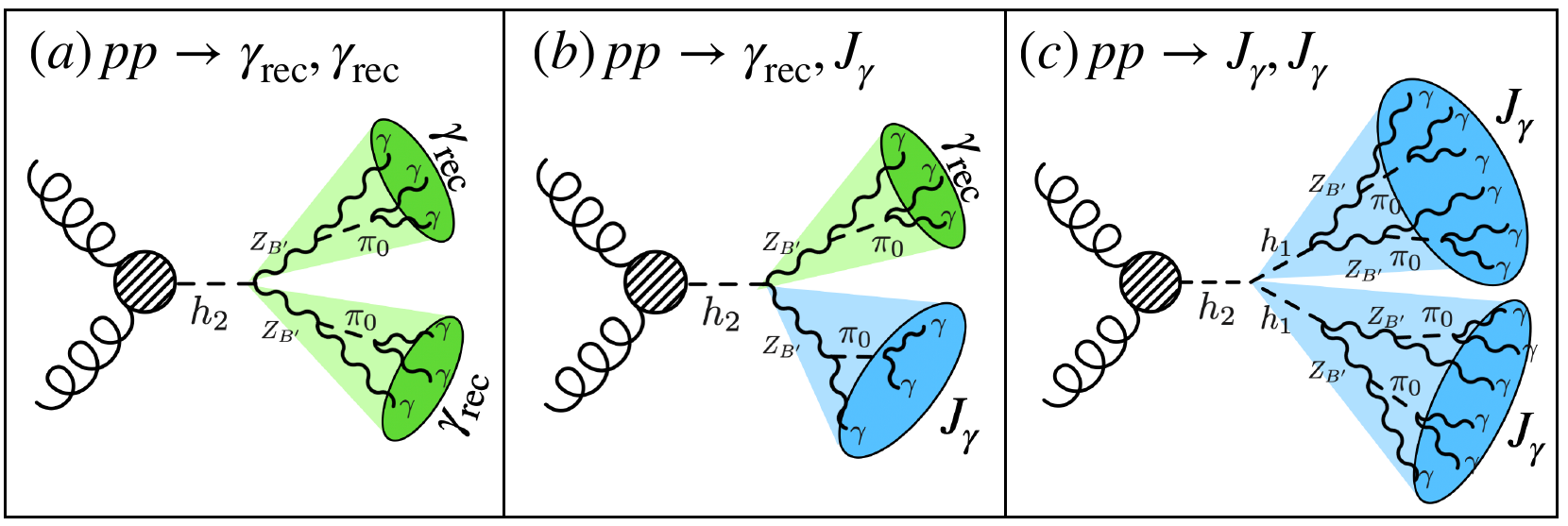}
\caption{
Signal signatures at the detector: (a) Two reconstructed photons, $ \gamma_{\text{rec}}\gamma_{\text{rec}} $ (left panel); (b) A reconstructed photon and a photon-jet, $ \gamma_{\text{rec}} J_{\gamma} $ (middle panel); (c) Two photon-jets, $ J_{\gamma}J_{\gamma} $ (right panel).
}
\label{fig:diagram}
\end{figure}

Based on the EM calorimeter structure  of ATLAS detector \cite{ATLAS:2018dfo}, the angular separation between two photons in the final state $ \Delta R_{\gamma\gamma}\lesssim 0.04 $ is approximately the same size as a standard single photon energy cluster.
In this case, existing triggers cannot distinguish a calorimeter energy deposit resulting from highly collimated photons from that of a single isolated photon.
Therefore, we use the Cambridge/Aachen (C/A) jet clustering algorithm~\cite{Dokshitzer:1997in,Wobisch:1998wt} with $ R = 0.04 $ to group these collimated photons. For those photons inside $ R = 0.04 $, they form a single reconstructed photon object, $ \gamma_{\text{rec}} $. However, for $ 0.04 < R < 0.4 $, we use the anti-$k_T$ jet clustering algorithm~\cite{Cacciari:2008gp} with the jet radius of $ R = 0.4 $ to group  collimated photons as is done in ordinary QCD jet analysis at the LHC. These photons are unable to pass the photon isolation criteria and thus are defined to form a photon jet, $ J_{\gamma} $\,\cite{Ellis:2012sd}. Accordingly, the signal signatures at the detector can be classified into 
\begin{enumerate}[label=(\alph*)]
\item Two reconstructed photons, $ \gamma_{\text{rec}}\gamma_{\text{rec}} $,  
\item A reconstructed photon and a photon-jet, $ \gamma_{\text{rec}} J_{\gamma} $,  
\item Two photon-jets, $ J_{\gamma}J_{\gamma} $,  
\end{enumerate} 
The three signal typologies are  shown in Fig.~\ref{fig:diagram} in which $ \gamma_{\text{rec}} $ or $ J_{\gamma} $ are defined according to the group of photons and the value of the radius parameter. Notice that each $ J_{\gamma} $ can include up to 3 collinear photons for the TP1 process and up to 6 collinear photons for the TP2 process, respectively.

\subsection{Events generation and detector simulation}\label{Sec:BPs}

To study these signal signatures, we establish a benchmark point (BP) for TP1 and four BPs for TP2, all with fixed values of $ \alpha_B = 10^{-5} $, $ M_{Z^{\prime}_B} = 0.4 $ GeV, and $ \sin\theta_h = 1.6\times 10^{-4} $. It's worth noting that for these BPs, $ BR(h_2\rightarrow h_1 h_1) $ is approximately $ 8.70\times 10^{-4} $, and $ BR(h_2\rightarrow Z^{\prime}_B Z^{\prime}_B) $ is roughly $ 4.35\times 10^{-4} $. 
 UFO model file is generated from \texttt{FeynRules}~\cite{Alloul:2013bka} for the simplified leptophobic $Z^{\prime}_B$.  \texttt{Madgraph5 aMC@NLO}~\cite{Alwall:2014hca,Frederix:2018nkq}  is used to compute the total cross section and event generation. Generated events are showered by \texttt{Pythia8}~\cite{Sjostrand:2007gs} to model the parton showering and hadronization. \texttt{Delphes3}~\cite{deFavereau:2013fsa} with modified ATLAS template is used for the fast detecor simulation. Moreover, we have implemented the photon conversion module in \texttt{Delphes3}, parameterized according to the conversion map of the CMS tracker as shown in Fig.~1 of Ref.~\cite{Selvaggi:2016ydq}. Note that the magnetic field in the detector is also simulated in the photon conversions module, which has been improved from the simulation in Ref.~\cite{Ellis:2012zp}.\texttt{FastJets}~\cite{Cacciari:2011ma} is used to cluster collimated photons in the final state. The basic photon isolation criteria in Delphes3 and the photon isolation criteria used by ATLAS~\cite{ATLAS:2018dfo}\footnote{Details about the ATLAS photon isolation criteria are provided in Sec.~\ref{Sec:Simulation}.} are applied, allowing for the distinction between $ \gamma_{\text{rec}} $ and $ J_{\gamma} $.

The results are summarized in Table~\ref{Tab:BP-2}, with the average number of reconstructed photons denoted as $ \langle N_{\gamma}\rangle $. The results without the photon conversion simulation are also shown in brackets for comparison. We find that the effect of photon conversions appears limited due to the following two main reasons: (1) The multiple collimated photons from the Higgs boson via gluon fusion are not energetic enough, resulting in $e^+ e^-$ pairs that are too soft; and (2) soft objects generated from Pythia8 simulations, such as multiparton interactions (MPI), have already significantly affected photon isolation. Additionally, we note that the photon transverse shower is not simulated in this analysis. In Refs.~\cite{ATLAS:2018dfo,Ellis:2012zp}, the photon transverse shower is considered a powerful variable for distinguishing between photons and QCD jets. 
However, since the current version of \texttt{Delphes3} for fast detector simulation does not include the simulation of the photon transverse shower, we did not include it in our analysis\footnote{Additionally, simulating the Moliere radius~\cite{Ellis:2012zp} could further improve the accurate identification of the number of photons within the photon jet, but this is beyond the scope of this work.}.
For TP1, the fraction of each classification is approximately one-third while TP2--BP1 and TP2--BP2 exhibit sizable probabilities for $ \gamma_{\text{rec}} J_{\gamma} $. In contrast, TP2--BP3 and TP2--BP4 are primarily classified as $ J_{\gamma}J_{\gamma} $. 
In the following, a detailed analysis is conducted, particularly regarding the $ \gamma_{\text{rec}} J_{\gamma} $ and $ J_{\gamma} J_{\gamma} $ signatures.
 \begin{table}[h!]
\centering
  \begin{tabular}{l r r r r r}
   \hline
BPs & $ M_{h_1} $ (GeV) & $\gamma_{\text{rec}}\gamma_{\text{rec}}$ & $\gamma_{\text{rec}} J_{\gamma}$ & $J_{\gamma}J_{\gamma}$ & $ \langle N_{\gamma}\rangle $ \\ \hline 
TP1 & $-$ & $27.51\%$ ($32.84\%$) & $36.36\%$ ($35.11\%$) & $36.13\%$ ($32.05\%$) & $0.91$ ($1.01$) \\ \hline
TP2--BP1 & $1.0$ & $4.44\%$ ($5.53\%$) & $30.59\%$ ($34.58\%$) & $64.97\%$ ($59.89\%$) & $0.39$ ($0.46$) \\ \hline
TP2--BP2 & $1.2$ & $2.69\%$ ($2.94\%$) & $23.36\%$ ($27.11\%$) & $73.95\%$ ($69.95\%$) & $0.29$ ($0.33$) \\ \hline
TP2--BP3 & $1.4$ & $1.12\%$ ($1.36\%$) & $17.52\%$ ($19.50\%$) & $81.36\%$ ($79.14\%$) & $0.20$ ($0.22$) \\ \hline
TP2--BP4 & $1.6$ & $0.59\%$ ($0.77\%$) & $13.06\%$ ($14.05\%$) & $86.35\%$ ($85.18\%$) & $0.14$ ($0.16$) \\ \hline
    \end{tabular}
    \caption{\small 
Branching fractions for $ \gamma_{\text{rec}}\gamma_{\text{rec}} $, $ \gamma_{\text{rec}} J_{\gamma} $, $ J_{\gamma}J_{\gamma} $ and the number of reconstructed photons, $ \langle N_{\gamma}\rangle $ for the five BPs with fixed $ \alpha_B = 10^{-5} $, $ M_{Z^{\prime}_B} = 0.4 $ GeV and $ \sin\theta_h = 1.6\times 10^{-4} $. The numbers in brackets here are without the photon conversion simulation.}
\label{Tab:BP-2}
\end{table}



\subsection{Preselection and event reconstruction}\label{Sec:Simulation}

In this subsection, we begin by examining the impact of the $ \gamma_{\text{rec}}\gamma_{\text{rec}} $ signature on the measurement of the SM-like Higgs boson to diphoton. Subsequently, we provide a detailed signal-to-background analysis and the corresponding significance for TP1 and TP2, as presented in Table~\ref{Tab:BP-2}, with a focus on the $ \gamma_{\text{rec}} J_{\gamma} $ and $ J_{\gamma} J_{\gamma} $ signatures.


In Table~\ref{Tab:BP-2}, when the final state with a pair of reconstructed photons cannot be distinguished from two isolated single photons in the ATLAS EM calorimeter, the signal can mimic the SM-like Higgs boson decay to diphoton~\cite{ATLAS:2012soa,Draper:2012xt}. According to Ref.~\cite{ATLAS:2020pvn}, the observed (SM expected) production rate for inclusive SM-like Higgs boson to diphoton is $ (\sigma\times B_{\gamma\gamma})_{\text{obs}}=127\pm 10 $ fb ($ (\sigma\times B_{\gamma\gamma})_{\text{exp}}=116\pm 5 $ fb). We estimate the contributions from TP1 and TP2 to the SM-like Higgs boson to diphoton production rate as follows: First, we adopt the total decay width of $ h_2 $ in Eq.~(\ref{eq:h2width-1}) to compute $ BR(h_2\rightarrow\gamma\gamma) $. Then, we rescale $ (\sigma\times B_{\gamma\gamma})_{\text{exp}} $ in our simplified leptophobic $ Z^{\prime}_B $ model accordingly. The total cross section, $ (\sigma\times B_{\gamma\gamma})_{\text{BSM}}$,  can be estimated as
\begin{equation}
(\sigma\times B_{\gamma\gamma})_{\text{BSM}}\equiv\sigma \cos^2\theta_h\times\left( BR(h_2\rightarrow Z^{\prime}Z^{\prime})\epsilon^{\text{iso}}_{\text{TP1}}+BR(h_2\rightarrow h_1 h_1)\epsilon^{\text{iso}}_{\text{TP2}}\right)\times \left(\frac{\epsilon^{\text{ID}}_{\text{BSM}}}{\epsilon^{\text{ID}}_{\text{SM}}}\right)^2\,,
\end{equation}
where $\sigma = 48.52$ pb is the Higgs boson production via gluon fuison at $\sqrt{s} = 13$ TeV with N$^3$LO QCD corrections and NLO electroweak corrections~\cite{ATLAS:2020pvn}.
$ \epsilon^{\text{iso}} $ represents the efficiency for collimated photons to satisfy the photon isolation criteria by ATLAS~\cite{ATLAS:2018dfo}; more details about this criteria discussed later. The $ \epsilon^{\text{ID}} $ follows Eq.~(1) in Ref.~\cite{ATLAS:2012soa}, where "ID" refers to photon identification based on various shower shape variables in different calorimeter layers. According to Fig.~7 in Ref.~\cite{ATLAS:2012soa}, we estimate $ \epsilon^{\text{ID}}_{\text{BSM}}/\epsilon^{\text{ID}}_{\text{SM}}\sim 0.7 $ for this study. Although, $ \epsilon^{\text{ID}}_{\text{BSM}} $ for TP2 should be smaller than the estimated value, the exact value can still computed using the \texttt{GEANT4} detector simulation~\cite{GEANT4:2002zbu}, which is beyond the scope of this work.

We first define the combinations as follows: ``TP1+TP2-BP1'' as Scenario I, ``TP1+TP2-BP2'' as Scenario II, ``TP1+TP2-BP3'' as Scenario III, and ``TP1+TP2-BP4'' as Scenario IV. Next, we present $ \epsilon^{\text{iso}}_{\text{TP1}} $, $ \epsilon^{\text{iso}}_{\text{TP2}} $, and $ (\sigma\times B_{\gamma\gamma})_{\text{BSM}} $ based on five BPs in Table~\ref{Tab:BP-2} for these four scenarios: 
\begin{itemize}
\item Scenario I: $\epsilon^{\text{iso}}_{\text{TP1}}=27.51\%$, $ \epsilon^{\text{iso}}_{\text{TP2-BP1}}=4.44\%$, $ (\sigma\times B_{\gamma\gamma})_{\text{BSM}} = 3.76$ fb;
\item Scenario II: $\epsilon^{\text{iso}}_{\text{TP1}}=27.51\%$, $ \epsilon^{\text{iso}}_{\text{TP2-BP2}}=2.69\%$, $ (\sigma\times B_{\gamma\gamma})_{\text{BSM}} = 3.40$ fb;
\item Scenario III: $\epsilon^{\text{iso}}_{\text{TP1}}=27.51\%$, $ \epsilon^{\text{iso}}_{\text{TP2-BP3}}=1.12\%$, $ (\sigma\times B_{\gamma\gamma})_{\text{BSM}} = 3.08$ fb;
\item Scenario IV: $\epsilon^{\text{iso}}_{\text{TP1}}=27.51\%$, $ \epsilon^{\text{iso}}_{\text{TP2-BP4}}=0.59\%$, $ (\sigma\times B_{\gamma\gamma})_{\text{BSM}} = 2.97$ fb.
\end{itemize}

We proceed with a detailed analysis of the $ \gamma_{\text{rec}} J_{\gamma} $ and $ J_{\gamma} J_{\gamma} $ signatures at the LHC. The primary sources of SM background for these two signal signatures originate from
\begin{align}
& p p\rightarrow\gamma\gamma, \quad p p\rightarrow\gamma j, \quad
p p\rightarrow jj.
\end{align}
Anti-$k_T$ algorithm is used to cluster calorimeter cells into jets with $ R = 0.4$.
We adhere closely to the photon isolation criteria outlined in Ref.~\cite{ATLAS:2018dfo}. Reconstructed photons must be isolated from other calorimeter energy deposits and nearby charged tracks not associated with the photon. The calorimeter isolation variable, denoted as $ E^{\text{iso}}_T $, is defined as the sum of energy deposits in the calorimeter within a cone of size $ \Delta R = 0.4 $ centered around the barycenter of the photon cluster (excluding the energy contributed by the photon cluster itself), minus $ 0.022\times E_T $. We require the calorimeter isolation variable to satisfy $ E^{\text{iso}}_T < 2.45 $ GeV. The track isolation variable, denoted as $ p^{\text{iso}}_T $, is the scalar sum of the transverse momenta of charged tracks not associated with the photon within a cone of size $ \Delta R = 0.2 $ centered around the barycenter of the photon cluster. It must satisfy $ p^{\text{iso}}_T < 0.05\times E_T $. A photon candidate is defined as a reconstructed photon, $\gamma_{\text{rec}}$, if it satisfies the following criteria: $ P^{\gamma}_T > 20 $ GeV, $ | \eta_{\gamma} | < 1.37 $ or $ 1.52 < | \eta_{\gamma} | < 2.37 $, and meets the aforementioned photon isolation criteria. After particle reconstruction, we define a photon-jet, $J_{\gamma}$, which must fulfill the following three conditions:
\begin{equation}
P^{J_{\gamma}}_T > 20 \text{ GeV}\quad \text{and}\quad | \eta_{J_{\gamma}} | < 2.5,
\end{equation}
\begin{equation}
\theta_{J_{\gamma}}=\frac{E_{J_{\gamma},\text{HCAL}}}{E_{J_{\gamma}}}\leqslant 0.1\,,
\end{equation}
\begin{equation}
\nu_{J_{\gamma}}=0\quad\text{with}\quad p_T > 2\text{ GeV},
\end{equation}
where $ \theta_{J_{\gamma}} $ represents the hadronic energy fraction, and $ \nu_{J_{\gamma}} $ denotes the number of charged particles within $J_{\gamma}$.

We further apply the following baseline event selections to identify signal events with the $ \gamma_{\text{rec}} J_{\gamma} $ signature:  
\begin{align}
& P_T(\gamma_{\text{rec1}}), P_T(J_{\gamma 1}) > 0.35 \ m_{\gamma_{\text{rec1}}J_{\gamma 1}}, 
\nonumber \\ & 
120 \text{ GeV} < m_{\gamma_{\text{rec1}}J_{\gamma 1}} < 130 \text{ GeV},
\nonumber \\ & 
|\eta_{\gamma_{\text{rec1}}}-\eta_{J_{\gamma 1}}| < 2.5,  
\end{align}
where the subscript label corresponds to the leading momentum photon. Reconstructed photon and the photon-jet are required to be highly boosted, with their invariant mass distributions mainly falling within the Higgs boson mass window. Similarly, we apply the following baseline event selections to identify signal events with the $ J_{\gamma} J_{\gamma} $ signature: 
\begin{align}
& P_T(J_{\gamma 1}) > 0.4 m_{J_{\gamma 1}J_{\gamma 2}}\quad \text{and}\quad P_T(J_{\gamma 2}) > 0.3 m_{J_{\gamma 1}J_{\gamma 2}}, 
\nonumber \\ & 
110 \text{ GeV} < m_{J_{\gamma 1}J_{\gamma 2}} < 140 \text{ GeV},
\nonumber \\ & 
|\eta_{J_{\gamma 1}}-\eta_{J_{\gamma 2}}| < 2.5.  
\end{align}
Furthermore, both photon-jets need to be highly boosted, enabling the reconstruction of the Higgs boson resonance peak from their invariant mass. The pileup effect is not considered in this study because it can be mitigated using traditional strategies, such as area subtraction~\cite{Cacciari:2008gn} and charged hadron subtraction~\cite{CMS:2014ata}. Additionally, modern machine learning algorithms like PUMML~\cite{Komiske:2017ubm} and others~\cite{ArjonaMartinez:2018eah, Berta:2019hnj, Maier:2021ymx} can effectively eliminate the pileup effect.
\section{Transformer network}\label{Sec:network}
Transformer models, introduced by Ref.~\cite{vaswani2017attention}, have revolutionized the field of deep learning and introduced in HEP in Refs.~\cite{Qu:2022mxj,Finke:2023veq,Hammad:2023sbd,He:2023cfc,Hammad:2024cae} . Unlike traditional sequence-based models, transformers utilize multi-heads self-attention mechanism to process input data in parallel, making them highly efficient for capturing long-range dependencies. This inherent capability makes transformers promising candidates for analyzing complex and interconnected systems, such as particle clouds. Adapting transformer models to analyze particle clouds involves encoding the spatial and temporal dynamics inherent in event collisions. Representing each particle as an input token, transformers can effectively capture interactions and correlations between particles. Moreover,  self-attention mechanism allows the model to weigh the significance of each particle's contribution to the overall cloud behavior. In this manner, transformer model can capture the important features shared by each individual particle to the whole cloud allowing for better classification performance between different signal and background events.  We point out to  previous  photon jets analysis using MLP, CNN and GNN \cite{Ren:2021prq,Wang:2023pqx,Ai:2024mkl}, which utilize image or graph-based input datasets. An improved representation for these analyses is an unordered, permutation-invariant set of particles, known as a particle cloud. This representation retains all the benefits of particle-based models, especially the flexibility to incorporate arbitrary features for each particle. Similar to the point cloud representation used in computer vision, particle clouds offer a robust framework for modeling jets.
Unlike graph or image representations that require particles to be sorted and thus implicitly depend on particle order, particle clouds impose no such order. Since particles within a jet lack intrinsic order, any manually imposed arrangement may be suboptimal and negatively impact performance. Additionally, particle clouds do not require preprocessing steps, which are essential for image and graph-based datasets and can cause networks to heavily rely on these transformations.
Traditional networks such as MLPs, CNNs, or basic GNNs are not permutation invariant and therefore cannot effectively analyze particle cloud datasets. In contrast, the Transformer network is permutation invariant and well-suited for particle cloud analysis. Additionally,  ATLAS experiment  utilize the photon transverse energy deposited in the ECAL to distinguish photons and misidentified hadronic jets \cite{ATLAS:2015rsn}. This variable has a discriminating validity as  the energy clusters deposited by photons in the EM calorimeter tend to be narrower in the transverse direction than those deposited by jets.  

In this paper, we employ two distinct networks utilizing attention-based transformer layers for TP1 and TP2 analysis. Both networks are designed to comprehensively analyze the event information encoded by the substructure of the photon-jet and reconstructed high-level kinematics. Since the information extracted from each benchmark point exhibits different structures, which cannot be effectively processed by a single network, we adapt multimodal networks comprising transformer and MLP layers for efficient modality learning.  

\subsection{Self-attention}
Self-attention mechanism leverages the relationships between particles within a cloud to compute attention scores, which determine the importance of each particle's features. By aggregating information from neighboring particles, self-attention enables the model to capture complex patterns and dependencies within the particle cloud, facilitating various tasks such as event classification and jet tagging. In this work we use the transformer encoder to extract the substructure information of the  photon-jet. Accordingly, we construct the input to the transformer layers as unordered sequence of particles inside the jet with a representation of
\begin{equation}
X = x_1^i,x_2^i \ \cdots \ x_n^i \,,
\end{equation}
where $i$ is the number of the corresponding features of each particle inside the jet and $n$ is the maximum number of included particles. To compute the attention score,  input representation is divided into three sets of vectors, Query (Q), Key (K) and Value (V). These vectors are obtained by linear projection of a fully connected layer to assign a learnable weight matrices as
\begin{equation}
Q = X W^Q, \hspace{4mm} K = X W^K, \hspace{4mm} V = X W^V \,, 
\end{equation}
where $W$ are the learnable weight matrices assigned to each vector via the linear projection.

Attention score can be obtained as scaled dot product of the three vectors as 
\begin{equation}
A = \frac{\text{exp}(Q.K^T/\sqrt{d})}{\sum\text{exp}(Q.K^T/\sqrt{d})}\,,
\end{equation}
where the sum is over the $K$ index and $d$ is the length of the input data\footnote{Normalizing the attention score is important to avoid the divergence of the dot product.}. Attention score is a matrix with dimensions equal to the number of the input particle tokens and  provide a flexible mechanism for modeling dependencies and relationships between particles in a sequence. By assigning different weights to each particle  based on its relevance to others, attention mechanisms can adaptively focus on the most salient information for the classification task. Moreover, it  provides a degree of interpretability by indicating which particles  in the input sequence are most influential for making predictions. The output of a single attention head is obtained as the dot product between the attention score and the Value vector, $\mathcal{Z} = A\cdot V$. The output of the attention head ensures the permutation invariance over the particles in the dataset by summing all the particle token via the matrix multiplication process.   
In order to speed up the computation and increase the rich of the network to capture more relevant information multi-heads can be used in parallel while their outputs can be stacked as
\begin{equation}
\widetilde{X} = \text{concatenate} \left(\mathcal{Z}_1, \mathcal{Z}_2 \ \cdots \ \mathcal{Z}_k    \right)\times W \,, 
\end{equation}   
where $k$ is the number of the parallel attention heads and $W$ is a normalizing matrix to keep the output dimensions as same as the inputs.  The output is then added to the input, via an addition wise layer, to weigh each of the particle tokens according to its computed attention. Exploiting the fact that transformer layer preserve the dimensions of the input dataset allows to stack many layers for better network  expressivity.
\subsection{Networks structure}
\label{subsec:network_structure}
\begin{figure}[t!]\centering
\includegraphics[width=0.98\textwidth]{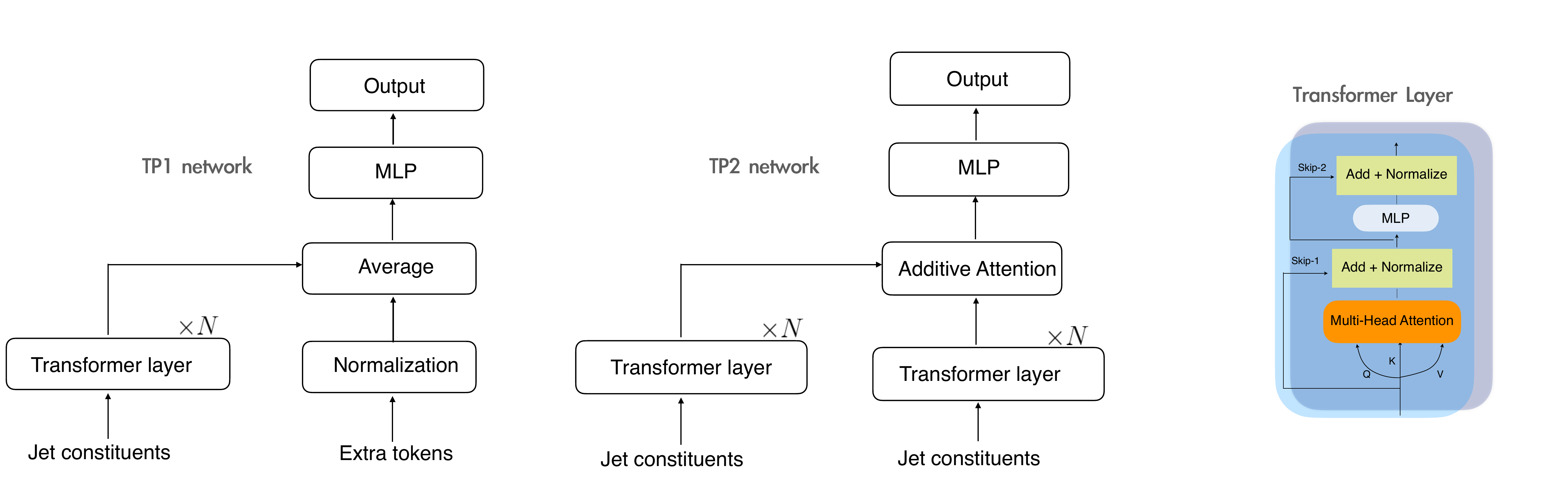}
\caption{ Left: Network architecture used for TP1 analysis. Middle: Network architecture used for TP2 analysis. Right: structure of the attention-based transformer layer used in the two networks. N indicates the number of repeated transformer layer. }
\label{fig:network}
\end{figure}

To fully leverage the capabilities of the two discussed channels\,$\left(\gamma_{\rm rec} J_{\gamma}, \,J_{\gamma}J_{\gamma}\right)$, we implement two distinct configurations of the transformer encoder with self-attention heads, as presented in Fig.~\ref{fig:network}. For the analysis of TP1, where we encounter a single photon-jet and one reconstructed photon, we tailor a transformer model with multi-head self-attention to analyze the constituents of the photon-jet, augmented with additional distributions  representing the reconstructed isolated photon kinematics.

The network is fed with two input datasets: the photon-jet contents, presented as a sequence of particles in a $(50,12)$ dimensional space, analyzed by transformer layers, and a supplementary dataset comprising five distributions, processed by a Fully Connected (FC) layer. The transformer layer is iterated three times, each with seven attention heads. The output from the attention heads is then integrated with the original input data via a skip connection layer.  The output of the final transformer layer is flattened and combined with the second dataset, then fed into an averaging layer. Subsequently, the output of the averaging layer is forwarded to an MLP with two FC layers of $128$ and $64$ neurons, respectively, utilizing Rectified Linear (ReLU) activation functions. Dropout layers, with dropout rate of $20\%$, follow each FC layer of the MLP, and the final output is directed to an FC layer with two neurons and softmax activation.

The second network used for TP2 analysis, consists of twins transformer encoders utilized to analyze the leading and subleading  photon-jet constituents.  
The network takes two separate data sets for the leading and subleading jet constituents as input with dimensions of $(50,12)$, processed through two distinct transformer layers. Each transformer layer is repeated three times and has the same structure as the transformer layer in TP1 network. In order to efficiently merge the output of the two transformers, we adopt an Additive attention layer \cite{Bahdanau:2014ghw}.  Additive attention dynamically computes attention weights based on the compatibility between the output of the two transformers, allowing the model to focus on relevant parts of both of them.
Compatibility score between the output of the first transformer, $H_i$, and the output of the second transformer, $E_j$, is computed as
\begin{equation}
A_{i,j} = \upsilon^T \cdot \text{tanh} \left( W_1\cdot H_i + W_2\cdot E_j  \right)\,,
\label{eq:A}
\end{equation}
where $\upsilon$ is a learnable parameter vector, $i.e,$ attention vector. The  output of the additive attention layer is obtained as 

\begin{equation}
\mathcal{O}_i = \sum_j \text{softmax} \left( A_{i,j} \right)\cdot E_j\,.
\end{equation} 
The output vector capture the weighted combination of elements from the second data set for each element of the first data set, allowing the model to leverage information from both data sets to perform classification task.
This output then passes through a MLP comprising two fully connected layers with dimensions $128$ and $64$, employing the GELU activation function. Following each fully connected layer, a dropout layer with a dropout rate of $20\%$ is applied. The output is then passed to the output layer for classification. 

Output of the two networks  is a probability with dimension ($\mathcal{P}_{sig},\mathcal{P}_{bkg} $), with $\mathcal{P}$ ranges from $[0,1]$. For event classification if the $\mathcal{P}_{sig}>0.5$, the event is classified as signal like event and if $\mathcal{P}_{sig}<0.5$ the event is classified as most likely background event. 

\subsection{Data structure}\label{subsec:network_structure}
\begin{figure}[h!]\centering
\includegraphics[width=0.95\textwidth]{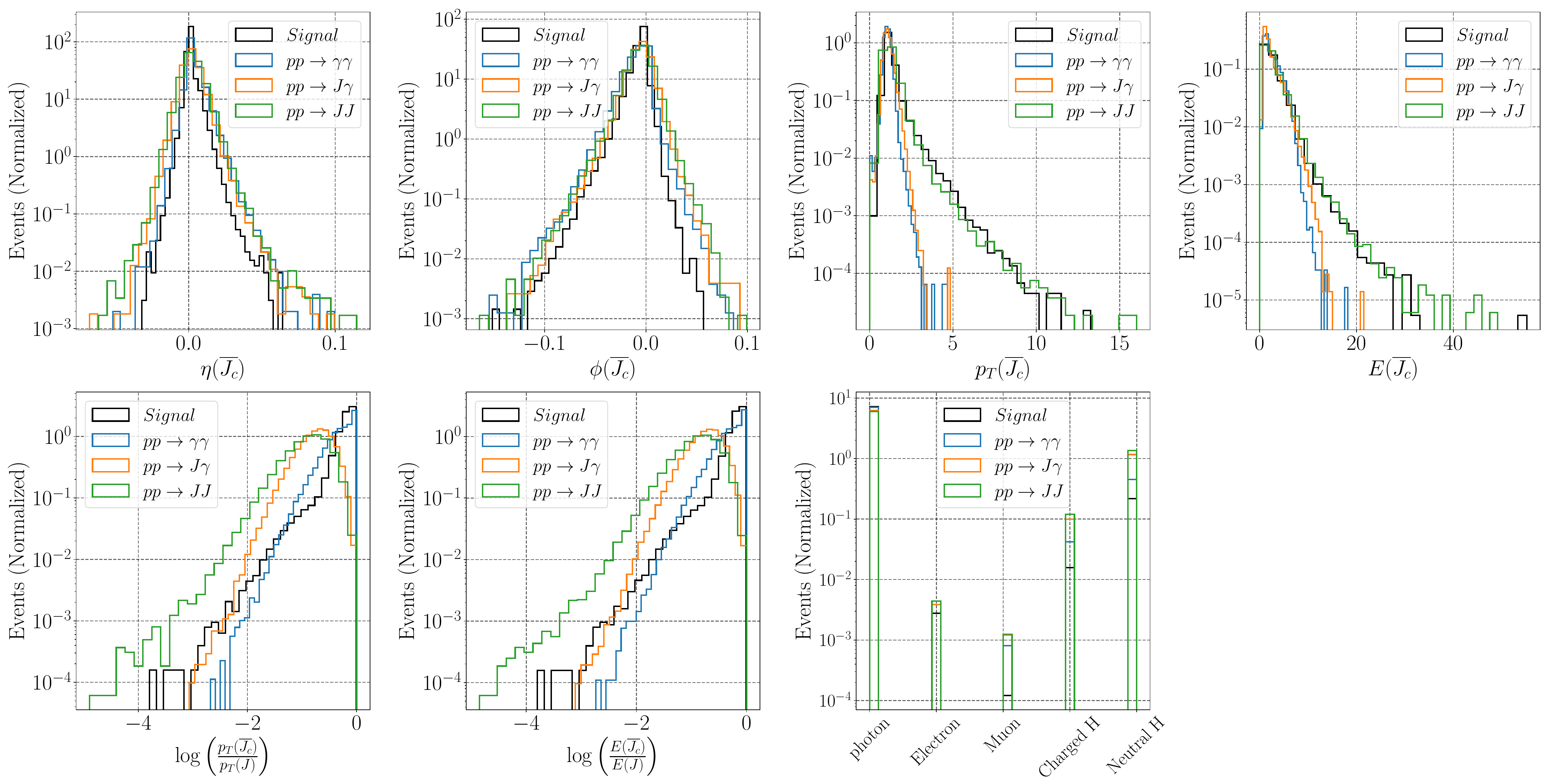}
\caption{Features of the leading photon-jet used as input to the  transformer layers. Each distribution is obtained by  averaging over the jet constituents in each event.}
\label{fig:inputs1}
\end{figure}
Configuring diverse data inputs for the network is achieved through particle clouds, emphasizing the permutation symmetry of inputs to yield a promising representation of jets. Initially, we pre-process the data sets for the leading photon-jet contents  by imposing a constraint of 50 constituents in each. The particles are arranged based on their transverse momentum in descending order. In instances where events comprise more than 50 constituents, only the top 50 are retained. For events with fewer constituents, the remaining positions are padded with zeros, ensuring conformity with the stipulated counts. An attention mask is used to notify the network with the padded events to not be further considered. In this case the results are independent on the prior hypothesized  number of constituents.  
We follow the variables input to the transformer layer from Ref.~\cite{Qu:2022mxj}, with the following:
\begin{enumerate}
\item $\Delta\eta_c = \eta_c-\eta_{jet}$, where $\eta_c (\eta_{jet})$ is the pseudorapidity of the jet constituents (reconstructed jet).
\item $\Delta\phi_c = \phi_c-\phi_{jet}$, where $\phi_c (\phi_{jet})$ is the azimuth angle of the jet constituents (reconstructed jet).  
\item $p_{T_c}$, transverse momentum of the jet constituents in GeV. 
\item $E_c$, energy of the jet constituents in GeV. 
\item $\log\left( \frac{p_{T_c}}{p_{T_{Jet}}} \right)$, transverse momentum of the jet constituents relative to the jet transverse momentum.
\item $\log\left( \frac{E_{c}}{E_{{Jet}}} \right)$, energy of the jet constituents relative to the jet energy.
\item $\Delta R_c = \sqrt{\Delta\eta^2_c+\Delta\phi^2_c}$, the angular distance of the jet constituents from the jet axis.
\item Particle ID for the jet constituents.
\end{enumerate}
The particle ID features are fed to the transformer model as a one-hot-encoding resulting extra 5 columns ending up with input features of dimensions $(50,12)$. 
To optimize the network discriminative accuracy, it is imperative to pre-process the jet contents, ensuring the manifestation of a multi-prong structure specific to signal events. For this purpose, the following transformations are applied before inputting the data into the network.  Features of the leading photon-jets which are used as input to the transformer layers are shown in Fig.\ref{fig:inputs1}. 
\begin{figure}[h!]\centering
\includegraphics[width=0.95\textwidth]{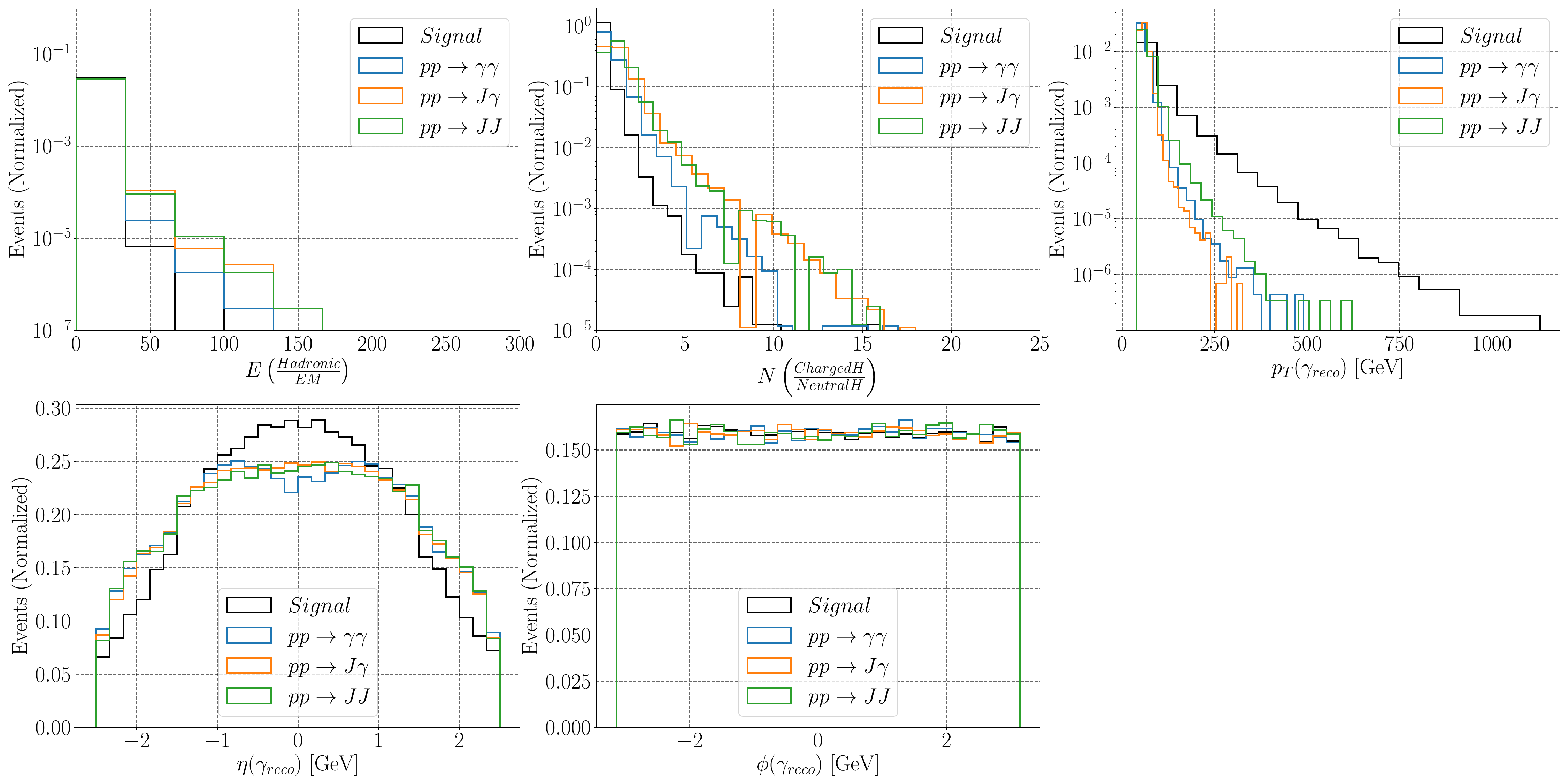}
\caption{Normalized events distributions for the extra tokens for the signal and backgrounds used  for TP1 process. With $E\left( \frac{Hadronic}{EM}\right)$ is the hadronic over electromagnetic energy of the leading jet and $N\left(\frac{ChargedH}{NeutralH} \right)$ is the numebr of the charged over the neutral particles inside the leading jet.}
\label{fig:inputs2}
\end{figure}

For TP1 process we consider full event information by investigating the kinematics of the reconstructed photon. Consequently, the second data set comprises five distributions representing the kinematics of the reconstructed photon  and  the photon-jet, as shown in Fig.\ref{fig:inputs2}.  For this purpose we consider the transverse momentum, pseudorapidity and azimuthal angle of the reconstructed photon with the relative energy deposit of the photon-jet in the different sub detectors. 
   
After preparing the data sets, we proceed to train the networks to recognize the intricate connections between the input data and their corresponding labels, which denote signal and background events. Signal events are labeled by Y = 1, while background events are labeled by Y = 0. To ensure the network's independence from the arrangement of signal and background events, we merge them into a single data set and randomize the order along with their assigned labels. During each epoch of network training (defined as a complete pass through the entire data sets), the network adjusts the weights of its neurons for each event through backward error propagation. The objective is to minimize the disparity between its predictions and the actual labels by converging to a global minimum of a defined loss function. This iterative process continues until the desired accuracy is achieved.

\section{Results}\label{Sec:results}
\begin{figure}[h!]\centering
\includegraphics[width=0.5\textwidth]{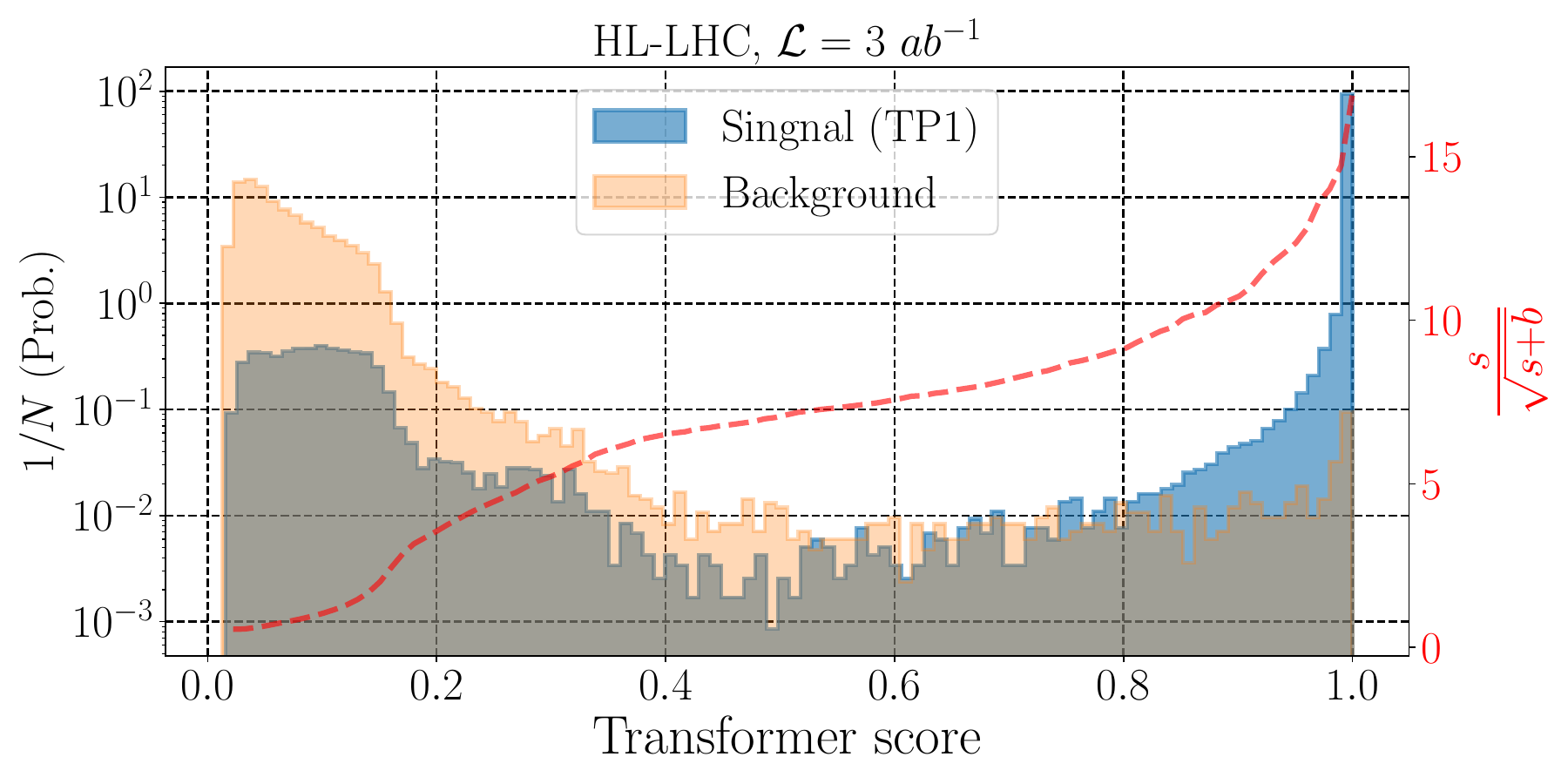}~~\includegraphics[width=0.5\textwidth]{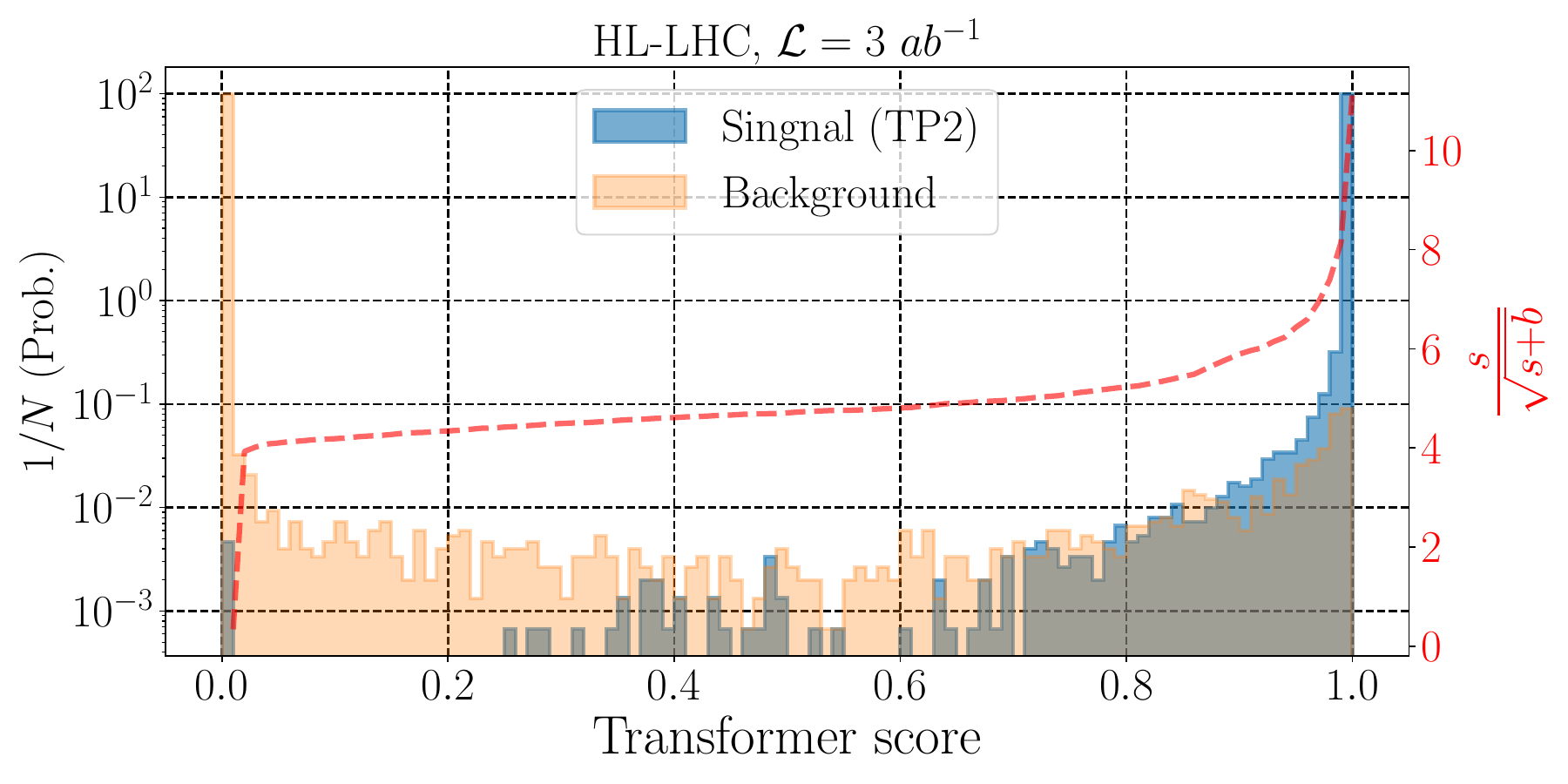}
\caption{Normalized output score of the transformer networks for TP1 (left) and TP2--BP2 (right). Blue and orange distributions present signal and background-like events, respectively. Red dashed line indicates the signal significance by setting a cut on the output score.  Signal significance is obtained at the HL-LHC. 
}
\label{fig:signeficance}
\end{figure}

In this section, we present the signal significance obtained by setting a cut on the networks output to maximize the signal to the background yield. Having the considered analysis for the two BPs, TP1 and TP2--BP2, at the $14$ TeV LHC with integrated luminosity of $3000$ fb$^{-1}$ (HL-LHC), the analysis results are presented in Fig.~\ref{fig:signeficance}. The discriminating power of each of the networks will be a measure of how well the signal and background characterised through their different features. Measuring the classification efficiency of each network Area Under the ROC (AUC) metric  is reported with $0.986$ and $0.998$ for TP1 and TP2--BP2, respectively. 

In Fig.~\ref{fig:signeficance}, the left panel illustrates the normalized output scores of the networks for TP1 analysis, while the right panel depicts TP2--BP2 analysis. The network output, represented as $(\mathcal{P}{\text{sig}}, 1 - \mathcal{P}{\text{bkg}})$, ranges from 0 to 1. Events with a discriminant value close to 1 are classified as signal-like events, depicted by the blue distribution, whereas those nearing 0 are categorized as background-like events, represented by the orange distribution. The intersection area between these two distributions measures the misclassification rate of each network. Signal significance is assessed by reweighting the network output for signal and background-like events based on the number of events remaining after preselection cuts. For TP1, there are $6375$ signal events and $5.6\times 10^9$ combined background events, while for TP2--BP2, there are $8700$ signal events and $2.5\times 10^{10}$  background events post the preselection cuts. Enhancing the signal to background yield is done by   maximizing the cut on the network output, which represented by the red dashed line.  The signal significance is approximated using the formula $\frac{S}{\sqrt{S+B}}$, where $S$ and $B$ represent the number of signal and background events, respectively. It shows a maximum value of $16 \sigma$ for TP1 and $11 \sigma$ for TP2--BP2.

\subsection{Projected limits on the model parameters}

Based on the baseline event selections and the transformer encoder analysis described above, we can extrapolate our results to establish future bounds on model parameters at the HL-LHC. Using $\frac{S}{\sqrt{S+B}} = 2$ for the $95\%$ Confidence Level (CL), we can determine the relevant cross-sections for signal and background processes from Fig.~\ref{fig:signeficance}. For TP1, the corresponding signal cross-section is $0.27$ fb, while the total SM background cross-section is $53.5$ fb. For TP2-BP2, the corresponding signal cross-section is $0.49$ fb, and the total SM background cross-section is $179$ fb. Although our analysis has significantly reduced the size of background cross-sections, they still exceed the signal cross-sections by two orders of magnitude.

\begin{figure}[h!]\centering
\includegraphics[width=0.48\textwidth]{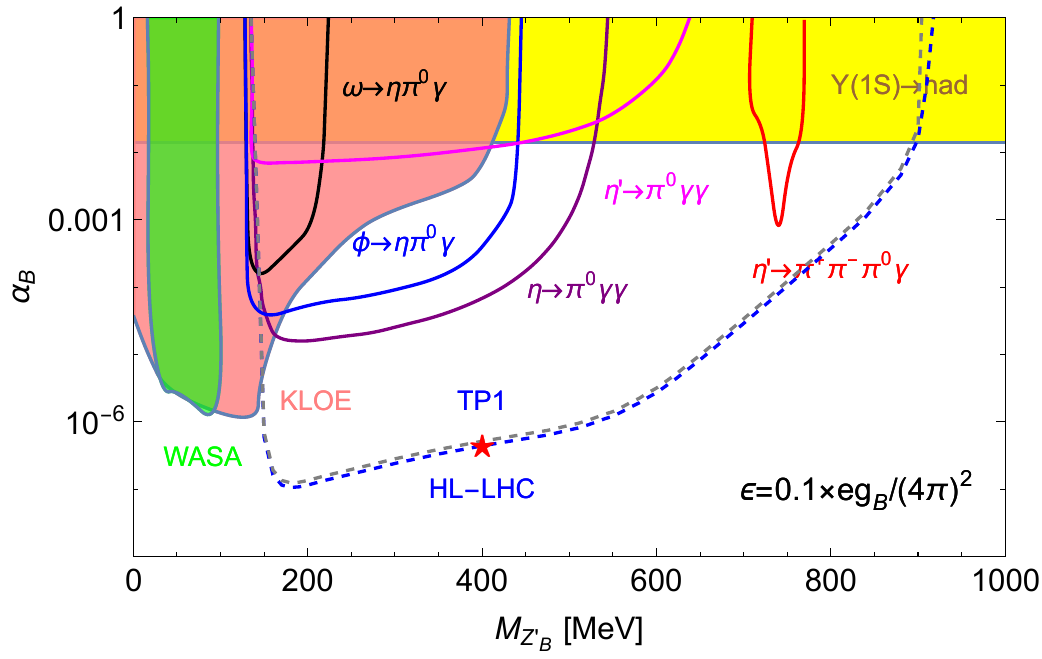}
\includegraphics[width=0.48\textwidth]{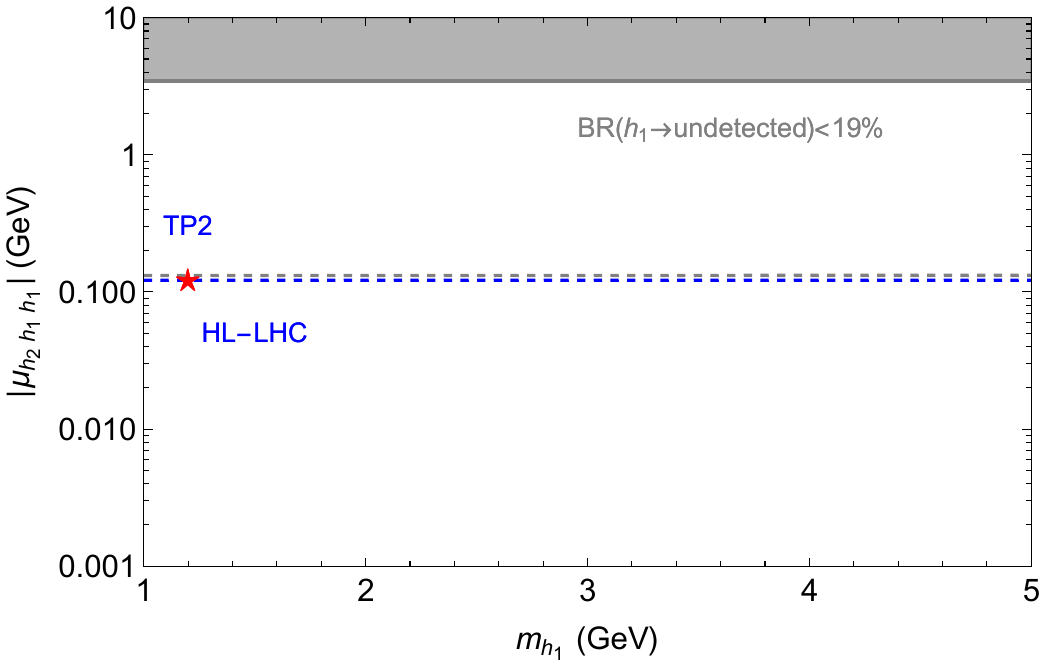}
\caption{Left: Projected  limits (blue dashed line) at $95\%$ CL on the plane $(M_{Z'_B}, \alpha_B)$ for TP1 at the HL-LHC. We fix $\sin\theta_H = 10^{-3}$ and label the red star as our BP. For comparison, the existing constraints from Fig.~\ref{fig:ZpBcons} are also shown. Right: The projection of limits (blue dashed line) at $95\%$ CL on the plane $(M_{h_1}, |\mu_{h_2 h_1 h_1}|)$ for TP2 at the HL-LHC. Here, $\sin\theta_H = 10^{-3}$, $\alpha_B = 10^{-5}$, and $M_{Z'_B} = 400$ MeV are fixed. Again, the red star is marked as our BP. The existing constraint for Higgs boson exotic decays from Eq.~\ref{eq:Higgs_exotic} is shown for comparison. The gray dashed lines correspond to including a $20\%$ systematic uncertainty of the SM background events in Eq.~\ref{eq:sig_Z}.}
\label{fig:summary}
\end{figure}

For TP1, $h_1$ is not involved. To simplify the analysis, we can exclude the contribution from $h_1$ and focus solely on the process $h_2\rightarrow Z'_B Z'_B$. According to Eqs.~(\ref{eq:h2width-1}), (\ref{eq:h2width-2}), and~(\ref{eq:h2ZpBZpB_width}), there are three free model parameters: $\sin\theta_H$, $\alpha_B$, and $M_{Z'_B}$. Assuming $\sin\theta_H = 10^{-3}$, future bounds on the plane $(M_{Z'_B}, \alpha_B)$ can be predicted, as indicated by the blue dashed line in the left panel of Fig.~\ref{fig:summary}. The red star represents our BP, which yielded $\alpha_B < 4.41\times 10^{-7}$ for $M_{Z'_B} = 400$ MeV at $95\%$ CL. It's important to note that only the BP (red star) is from our actual analysis; we assume that other $Z'_B$ masses (blue dashed line) have the same efficiency as the BP. Since the light $Z'_B$ is highly boosted from the Higgs boson decay, we expect the efficiency for other $Z'_B$ masses to be only mildly different from that of the BP. However, the dashed line rises rapidly as $M_{Z'_B}$ increases, mainly due to the shape of the branching ratio of $Z'_B\rightarrow\pi^0\gamma$ in Fig.~\ref{fig:ZpBcons}. Consequently, future bounds on $\alpha_B$ may weaken as the size of $\sin\theta_H$ decreases.

For TP2, both $h_1$ and $Z'_B$ are involved, leading to the simultaneous appearance of Higgs boson exotic decays, $h_2\rightarrow h_1 h_1$ and $h_2\rightarrow Z'_B Z'_B$. In this scenario, there are five free model parameters: $\sin\theta_H$, $\alpha_B$, $M_{Z'_B}$, $M_{h_1}$, and $|\mu_{h_2 h_1 h_1}|$. By fixing $\sin\theta_H = 10^{-3}$, $\alpha_B = 10^{-5}$, and $M_{Z'_B} = 400$ MeV, future bounds on the plane $(M_{h_1}, |\mu_{h_2 h_1 h_1}|)$ can be predicted, as indicated by the blue dashed line in the right panel of Fig.~\ref{fig:summary}. Here, the red star represents our BP, which yielded $|\mu_{h_2 h_1 h_1}| < 0.12$ GeV for $M_{h_1}=1.2$ GeV at $95\%$ CL. Similarly, except for the BP (red star), other $M_{h_1}$ values (blue dashed line) are assumed to have the same efficiency as the BP. The constraint for Higgs boson exotic decays from Eq.~\ref{eq:Higgs_exotic} is also added in Fig.~\ref{fig:summary} for comparison. Moreover, future bounds on $|\mu_{h_2 h_1 h_1}|$ can be weakened when one or both of $\sin\theta_H$ and $\alpha_B$ decrease.

Regarding the systematic uncertainties arising from QCD di-jet measurements, which may affect the reported classification performance, we consider a $20\%$ systematic uncertainty to estimate this effect. The formula for signal significance is modified accordingly \cite{LHCDarkMatterWorkingGroup:2018ufk,Antusch:2018bgr}: 
\begin{equation}
Z = \sqrt{2\cdot\left[(S+B)\cdot ln\left(\frac{(S+B)(B+\sigma^2_B)}{B^2 +(S+B)\sigma^2_B}\right)-\frac{B^2}{\sigma^2_B}\cdot ln\left(1+\frac{\sigma^2_B S}{B(B+\sigma^2_B)}\right)\right]} \,,
\label{eq:sig_Z}
\end{equation}
where $\sigma_B$ represents the systematic uncertainty of the SM background events $B$. In our analysis results, the background events $B$ are much larger than the signal events $S$, so the projected limits accounting for a $20\%$ systematic uncertainty on the model parameters only change slightly at $95\%$ CL, as shown by the gray lines in Fig.~\ref{fig:summary}.

\subsection{Decoding networks output}
Interpretable ML methods are crucial because they provide transparency and insights into how models make decisions, which is essential for trust, accountability, and debugging. Analyzing the network's latent space, in particular, is crucial because it reveals how the model internally represents and processes data, shedding light on the underlying structure and patterns it has learned. This understanding helps in diagnosing and correcting errors, enhancing model robustness, and ensuring that the model's decision-making process aligns with the physics of the considered process.

\begin{figure}[h!]\centering
\includegraphics[width=0.87\textwidth]{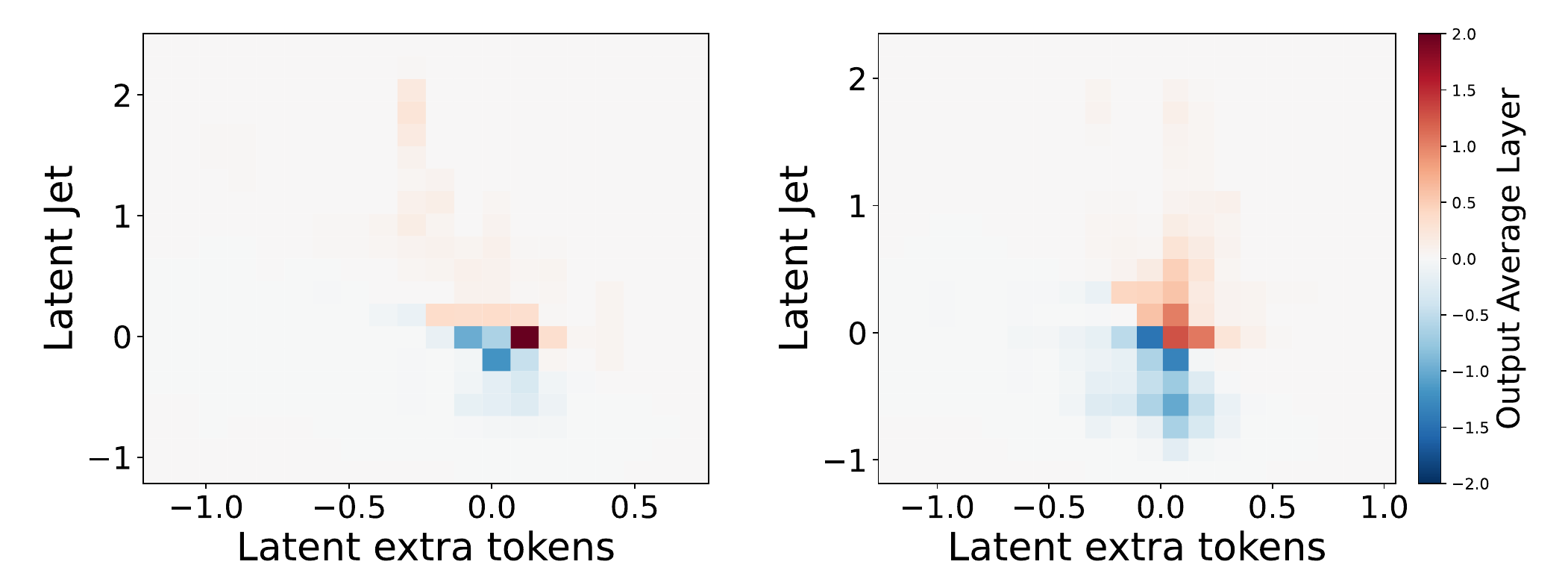}
\caption{
Latent space distribution of the jet constituents and extra tokens for $10^5$ test events  (left) and background events (right). Color bar indicates the output of the element-wise average layer.}
\label{fig:latent_1}
\end{figure}

After determining the signal significance of the two benchmark points by optimizing the cut on the networks' output, we proceed to delve into the network decision making process by conducting a numerical analysis of each network's flow. In Fig.~\ref{fig:latent_1}, we present the latent space distribution of both the transformer and MLP, utilized for additional tokens analysis, weighted by the output values of the element-wise average layer.

The latent space representation of the MLP indicates a degree of similarity between signal and background events, while the latent distribution of the transformer layers showcases a broader variation among background events compared to signal events. This disparity stems from the inherent structure of the photon-jet in TP1, characterized by a prong structure where the particles with the highest momenta are concentrated around the jet axis. In contrast, the QCD jet lacks such a prong structure, resulting in varied momenta of the particles within the jet distributed around the QCD jet axis.

\begin{figure}[h!]\centering
\includegraphics[width=0.87\textwidth]{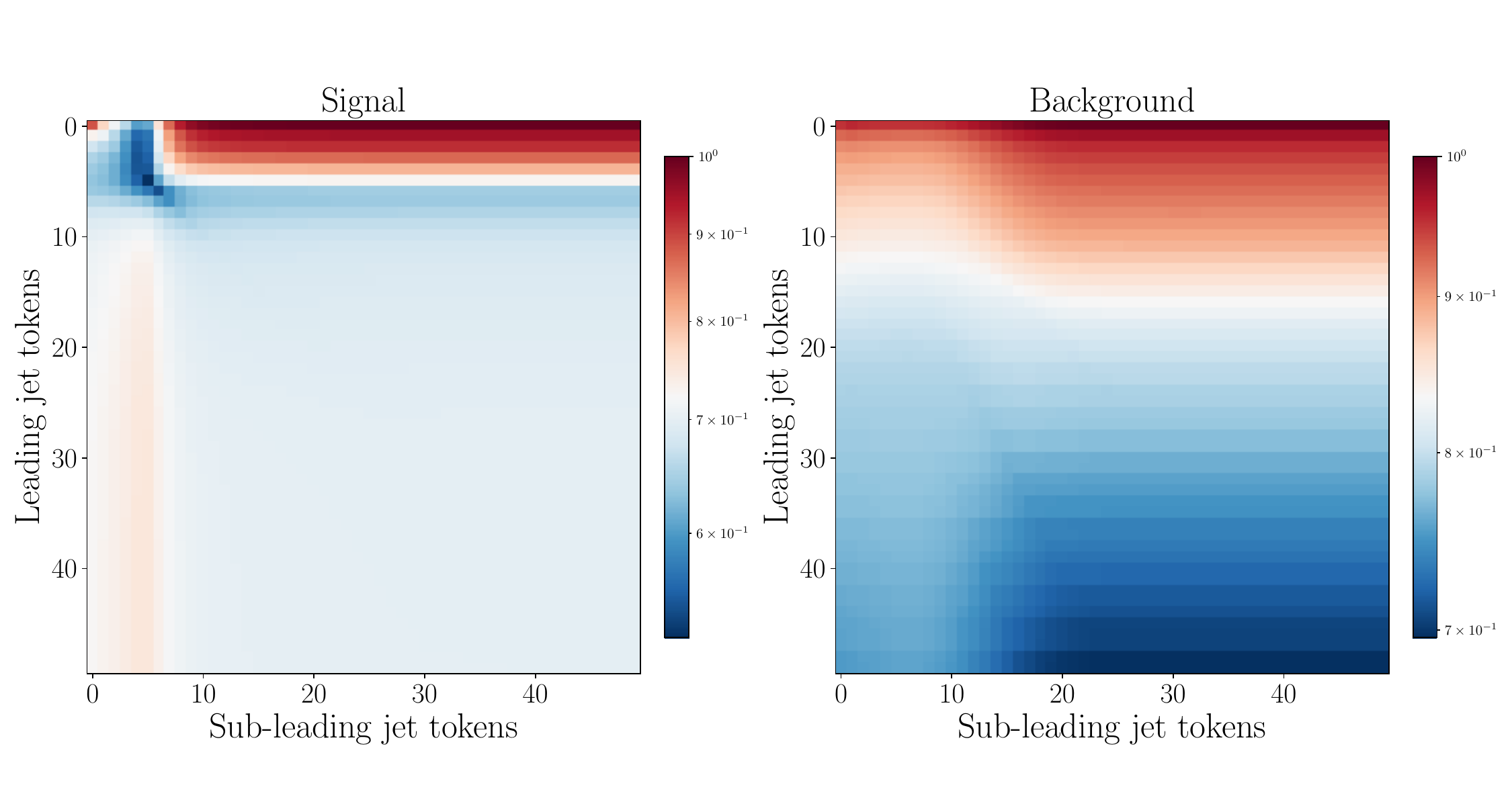}
\caption{Attention weights of the particle tokens for the leading and sub-leading jet for the signal events (left) and background events (right). The attention weights are normalized for the $10^5$ test images. Color bar indicates the value of the attention scores as in Eq.~(\ref{eq:A}).}
\label{fig:latent_2}
\end{figure}

For TP2, we investigate the network decision making by visualizing the attention score, $A_{i,j}$ in Eq.~(\ref{eq:A}). These scores have dimensions corresponding to the number of particle tokens in the leading and subleading  jets, i.e., $(50,50)$, offering insight into the relationships and dependencies between different particles in the respective datasets. Essentially, attention scores highlight which particle tokens from the two datasets, the leading and subleading jet constituents, play pivotal roles in the network's decision-making process.

Figure \ref{fig:latent_2} visualizes the attention scores, $A_{i,j}$, generated by the additive attention layer for $10^5$ signal events (left panel) and background events (right panel). Intriguingly, the network predominantly focuses on the first five highest momentum particles inside the leading and subleading  jet to identify signal events, whereas for background events, the network distributes high attention across a wider range of particle contents. We interpret this behavior as the network learning the distinct substructures of photon and QCD jets. 
Therefore, it makes the network robust against random fluctuations in the photon jet substructure caused by potential pileup effects.

\section{Conclusion}\label{Sec:Conclusion}

Although the SM has been so successful in a wide range of energy scales, there are a number of reasons to consider its extensions in various directions.
One possible extension would be $U(1)_B$, which may help to understand better the longevity of proton or the origin of global $U(1)_B$ symmetry of the SM. 
In this paper we considered a simplified model for $U(1)_B$ extensions of the SM, assuming all the exotic particles are heavy except 
for the $U(1)_B$ gauge boson $Z_B^{'}$ and the $U(1)_B$-breaking Higgs boson $h_B \simeq h_1$. 
Here $h_B \simeq h_1$ can be pair-produced in the SM Higgs decays, where $h_{\rm SM} (\simeq h_2)$ decays into $h_1 h_1$ if kinematically allowed. Additionally, if this channel is open, $h_1$ will mainly decay into a pair of $Z_B^{'}$. For a certain mass region of $Z_B^{'}$ ($m_{\pi^0} \lesssim m_{Z_B^{'}} \lesssim 620$ MeV), 
the main decay channel would be $Z_B^{'} \rightarrow \pi^0 \gamma \rightarrow ( \gamma \gamma ) \gamma$, and not $\pi^+ \pi^-$ as shown in Fig.~\ref{fig:ZpBBR}. 
This is because $Z_B^{'}$ is an iso-singlet vector meson, just like $\omega^0 (783)$ in ordinary QCD 
\footnote{$\rho^0 (770)$ is in the similar mass region as $Z_B^{'}$, and its main decay channel is $\rho^0 (770) \rightarrow \pi^+ \pi^-$, 
since it is an isovector ($I=1$) vector ($J=1$) meson. Note that $\rho^0 \rightarrow \pi^0 \pi^0$ is strictly forbidden by the total angular 
momentum conservation and spin-statistics theorem.}.
When $Z_B^{'}$ boson is boosted, its decay products will make a reconstructed photon or a photon-jet, unique signatures of our simplified $U(1)_B$ model 
with light $Z_B^{'}$ and $h_1$ bosons as shown in Fig.~\ref{fig:diagram}. One has to pay extra care for its study at high energy colliders, especially at the HL-LHC.  
Compared with previous photon-jet studies at the LHC, such signatures emerging from $Z^\prime_B$ decays have not been considered before, as far as we know. This motivates us to focus on these types of signatures. 

For this purpose, we adopted  advanced machine learning techniques, in particular the transformer encoder to identify the novel photon-jet signature and distinguish it from the SM background events.
Unraveling the distinctive characteristics of these typologies at the LHC presents a formidable challenge due to significant contamination from the QCD di-jet process within the signal events. To address this, and to enhance the signal-to-background ratio, we employ two ML methodologies centered on self-attention transformer encoders. Our investigation focuses on two specific processes: $h_2\to Z^\prime_B Z^\prime_B$ (TP1) and $h_2\to h_1 h_1$ (TP2), yielding $6$ and $12$ photons, respectively. Within these final states, we examine processes involving $\gamma_{reco}+J_\gamma$ and $J_\gamma+J_\gamma$. Recognizing that each final state encapsulates distinct event features, we employ two slightly different ML networks. The first network, analyze the $\gamma_{reco}+J_\gamma$ final state, comprises a transformer layer for analyzing photon-jet substructure, complemented by a MLP encoding the kinematics of reconstructed photons and photon-jets. The second network, focusing on the $J_\gamma+J_\gamma$ final state, incorporates two transformer layers to analyze the substructure of both photon-jets. These two network architectures are displayed in Fig.~\ref{fig:network}.

The structured architecture of these networks enables robust discrimination between signal and background events, achieving classification accuracies exceeding $98\%$, as shown in Fig.~\ref{fig:signeficance}. Using our baseline event selections and the transformer encoder analysis, we can forecast the future bounds of model parameters in this simplified leptophobic $ Z^{\prime}_B $ model at the HL-LHC. For TP1, fixing the mixing angle between $h_B$ and $h_{\rm SM}$ at $\sin\theta_H = 10^{-3}$, we found that the baryonic force fine structure constraint, $\alpha_B$, can be as low as $5\times 10^{-7}$ for $M_{Z'_B}\sim 400$ MeV at $95\%$ CL. For TP2, fixing both $\sin\theta_H = 10^{-3}$ and $\alpha_B = 10^{-5}$, the three-point interaction $h_2$-$h_1$-$h_1$ can be constrained to $|\mu_{h_2 h_1 h_1}| < 0.12$ GeV for $M_{h_1}\sim 1$ GeV at $95\%$ CL. Ultimately, our results suggest that probing photon-jet signatures remains feasible at the HL-LHC, even when accounting for $20\%$ of systematic uncertainties. Therefore, searching for Higgs boson exotic decays in energy frontier experiments can link the physics at the electroweak scale to the dark sector at the low energy scale and serve as a complementary probe to search for light $Z'_B$ and $h_1$ in intensity frontier experiments.

\section*{Acknowledgment} 
This work is supported in part by 
the KIAS Individual Grants under Grant No. PG021403 at Korea Institute for Advanced Study (P.K.), National Research Foundation of Korea NRF-2021R1A2C4002551 (M.\,P.\,), 
the National Natural Science Foundation of China (NNSFC) under grant No. 12335005 (C.-T. Lu). AH is funded by grant number 22H05113,
“Foundation of Machine Learning Physics”, Grant in
Aid for Transformative Research Areas and 22K03626,
Grant-in-Aid for Scientific Research (C).


\appendix

\section{Analytical formulae of light $ h_1 $ partial decay widths}\label{Sec:h1_decay}

In this appendix, we show analytical formulae of light $ h_1 $ partial decay widths used in our work~\cite{Chang:2013lfa,Liu:2014cma}. For $ M_{h1}\sim {\cal O}(1) $ GeV, the $ h_1 $ total width, $ \Gamma_{h_1} $, can be written as   
\begin{equation}
\Gamma_{h_1} = \cos^2\theta_h\widehat{\Gamma}(h_1\rightarrow Z^{\prime}_B Z^{\prime}_B)+\sin^2\theta_h\left(\widehat{\Gamma}(h_1\rightarrow f\overline{f})+\widehat{\Gamma}(h_1\rightarrow\gamma\gamma)\right)
\end{equation}
where $ f\overline{f} $ can be charged lepton pairs or light quarks to form meson pairs. These $ h_1 $ partial decay widths are
\begin{align}
& \widehat{\Gamma}(h_1\rightarrow Z^{\prime}_B Z^{\prime}_B) = \frac{g^2_B M^2_{Z^{\prime}_B}}{8\pi M_{h_1}}\sqrt{1-\frac{4M^2_{Z^{\prime}_B}}{M^2_{h_1}}}(3-\frac{M^2_{h_1}}{M^2_{Z^{\prime}_B}}+\frac{M^4_{h_1}}{4M^4_{Z^{\prime}_B}}), \\ &
\widehat{\Gamma}(h_1\rightarrow \ell^+\ell^-) = \frac{m^2_{\ell}M_{h_1}}{8\pi v^2_B}(1-\frac{4m^2_{\ell}}{M^2_{h_1}})^{\frac{3}{2}}, \\ &
\widehat{\Gamma}(h_1\rightarrow \pi^+\pi^-) = \frac{M^3_{h_1}}{324\pi v^2_B}\sqrt{1-\frac{4m^2_{\pi}}{M^2_{h_1}}}(1+\frac{11m^2_{\pi}}{2M^2_{h_1}})^2, \\ &
\widehat{\Gamma}(h_1\rightarrow \pi^0\pi^0) = \frac{1}{2}\times\widehat{\Gamma}(h_1\rightarrow \pi^+\pi^-), \\ &
\widehat{\Gamma}(h_1\rightarrow K^+ K^-) = \widehat{\Gamma}(h_1\rightarrow K^0\overline{K^0}) = \frac{M^3_{h_1}}{324\pi v^2_B}\sqrt{1-\frac{4m^2_K}{M^2_{h_1}}}(1+\frac{11m^2_K}{2M^2_{h_1}})^2, \\ &
\widehat{\Gamma}(h_1\rightarrow\eta\eta) = \frac{M^3_{h_1}}{648\pi v^2_B}\sqrt{1-\frac{4m^2_{\eta}}{M^2_{h_1}}}(1+\frac{11m^2_{\eta}}{2M^2_{h_1}})^2, \\ &
\widehat{\Gamma}(h_1\rightarrow\gamma\gamma) = \frac{\alpha^2_{\text{EM}}M^3_{h_1}}{256\pi^3 v^2_B}\lvert\sum_i N^i_c (Q^i)^2 F_{\frac{1}{2}}(\frac{4m^2_i}{M^2_{h_1}})+F_1(\frac{4m^2_W}{M^2_{h_1}})\rvert^2,
\end{align}
where $ \alpha_{\text{EM}} $ is the fine structure constant, $ N_c $ is the color factor and $ Q $ is the electric charge. The functions $ F_1(x), F_{\frac{1}{2}}(x) $ are
\begin{align}
& F_1(x) = 2+3x+3x(2-x)f(x), \\ &
F_{\frac{1}{2}}(x) = -2x[1+(1-x)f(x)], 
\end{align}
which
\begin{eqnarray}
f(x) = \left\{
       \begin{array}{lr}
   \left(\sin^{-1}\sqrt {1/x} \right)^2 , & \quad x \ge 1, 
   \\
    - \frac{1}{4}\left[\ln \left(\frac{{1 + \sqrt {1 - x} }}
      {{1 - \sqrt {1 - x} }}\right) - i \pi \right]^2 , & \quad x < 1. 
\end{array}
      \right.
\end{eqnarray}
Notice there is no reliable description of non-perturbative QCD effects can be presented in the mass region $ M_{h_1}\simeq 1.5-2.5 $ GeV~\cite{Bezrukov:2009yw}. Therefore, we only focus on $ M_{h_1}\lesssim 1.5 $ GeV in this appendix.

\section{Analytical formulae of light $ Z^{\prime}_B $ partial decay widths}\label{Sec:Zp_decay}
In this appendix, we collect analytical formulae of $ Z^{\prime}_B $ partial decay widths used in our work which closely follows Ref.~\cite{Tulin:2014tya}. The light $ Z^{\prime}_B $ total decay width, $\Gamma_{Z^{\prime}_B}$, can be written as 
\begin{align}
\Gamma_{Z^{\prime}_B} = & \Gamma(Z^{\prime}_B\rightarrow\pi^0\gamma)+\Gamma(Z^{\prime}_B\rightarrow\pi^+\pi^-\pi^0)+\Gamma(Z^{\prime}_B\rightarrow\pi^+\pi^-)+\Gamma(Z^{\prime}_B\rightarrow\eta\gamma) \nonumber  \\ & 
+ \Gamma(Z^{\prime}_B\rightarrow l^+l^-)
\end{align}
where
\begin{itemize}
\item $ \Gamma(Z^{\prime}_B\rightarrow\pi^0\gamma) $:
\begin{equation}
\Gamma(Z^{\prime}_B\rightarrow\pi^0\gamma) = \frac{\alpha_{B}\alpha_{\text{EM}}M^3_{Z^{\prime}_B}}{96\pi^3 f_{\pi}^2}(1-\frac{m^2_{\pi}}{M^2_{Z^{\prime}_B}})^3 \lvert F_{\omega}(M^2_{Z^{\prime}_B})\rvert^2
\end{equation}
where $ f_{\pi}\simeq 93 $ MeV is the pion decay constant and $ \alpha_{B}\equiv\frac{g^2_B}{4\pi} $. The form factor $ F_{\omega}(s) $ is given by $ F_{\omega}(s)\approx 1-s/m^2_{\omega}-i\Gamma_{\omega}/m_{\omega} $.

\item $ \Gamma(Z^{\prime}_B\rightarrow\pi^+\pi^-\pi^0) $:
\begin{equation}
\Gamma(Z^{\prime}_B\rightarrow\pi^+\pi^-\pi^0) = \frac{g^4_{\rho\pi\pi}\alpha_{B}M_{Z^{\prime}_B}}{192\pi^6 f_{\pi}^2}{\cal I}(M^2_{Z^{\prime}_B})\lvert F_{\omega}(M^2_{Z^{\prime}_B})\rvert^2
\end{equation}
where $ g_{\rho\pi\pi} $ is the $ \rho\pi\pi $ coupling and fixed to be $ g^2_{\rho\pi\pi}/(4\pi)\simeq 3.0 $ from the observed $ \rho\rightarrow\pi\pi $ decay rate. The details of the integral over phase space $ {\cal I}(M^2_{Z^{\prime}_B}) $ can be found in the Appendix of Ref.~\cite{Tulin:2014tya}.

\item $ \Gamma(Z^{\prime}_B\rightarrow\pi^+\pi^-) $:
\begin{equation}
\Gamma(Z^{\prime}_B\rightarrow\pi^+\pi^-) = \frac{\alpha_{\text{EM}}\epsilon^2 M_{Z^{\prime}_B}}{12}(1-\frac{4m^2_{\pi}}{M^2_{Z^{\prime}_B}})^{\frac{3}{2}}\lvert F_{\pi}(M^2_{Z^{\prime}_B})\rvert^2
\end{equation}
where
\begin{equation}
F_{\pi}(s)=F_{\rho}(s)[1+\frac{1+\delta}{3}\frac{\overline{\Pi}(s)}{s-m^2_{\omega}+im_{\omega}\Gamma_{\omega}}].
\end{equation}
If $ \delta = 0 $, then $ \overline{\Pi}(m^2_{\omega}) = -3500\pm 300 $ MeV$^2$. Due to the direct mixing between $ Z^{\prime}_B $ and $ \omega $ meson, $ \delta = 2g_B /(\epsilon e) $ is considered. For lack of a better understanding of $ \overline{\Pi}(s) $, we assume a constant value $ \overline{\Pi}(s)\approx\overline{\Pi}(m^2_{\omega}) $ here, although this assumption is not well justified. On the other hand, we take a simple Breit-Wigner form $ F_{\rho}(s)\approx 1-s/m^2_{\rho}-i\Gamma_{\rho}/m_{\rho} $ as an approximation.

\item $ Z^{\prime}_B\rightarrow\eta\gamma $:
\begin{equation}
\Gamma(Z^{\prime}_B\rightarrow\eta\gamma) = \frac{\alpha_{B}\alpha_{\text{EM}}M^3_{Z^{\prime}_B}}{96\pi^3 f_{\eta}^2}(1-\frac{m^2_{\eta}}{M^2_{Z^{\prime}_B}})^3 \lvert F_{\phi}(M^2_{Z^{\prime}_B})\rvert^2
\end{equation}
where the form factor $ F_{\phi}(s) $ is given by $ F_{\phi}(s)\approx 1-s/m^2_{\phi}-i\Gamma_{\phi}/m_{\phi} $.

\item $ Z^{\prime}_B\rightarrow \ell^+\ell^- $:
\begin{equation}
\Gamma(Z^{\prime}_B\rightarrow\ell^+\ell^-) = \frac{\alpha_{\text{EM}}\epsilon^2 M_{Z^{\prime}_B}}{3}(1+\frac{2m^2_{\ell}}{M^2_{Z^{\prime}_B}})\sqrt{1-\frac{4m^2_{\ell}}{M^2_{Z^{\prime}_B}}}.
\end{equation}
\end{itemize}


\end{document}